\documentclass{aa}  

\usepackage{hyperref}
\hypersetup{
    colorlinks = true,
    allcolors = {blue}
}
\usepackage{graphicx}
\usepackage{txfonts}
\usepackage{lipsum}

\newcommand{\mjup}{\mbox{~$M_{Jup}$}}
\newcommand{\prim}{\mbox{HR~8799}}

\begin{document} 

   \title{A search for a 5th planet around \prim\ using the star-hopping RDI technique at VLT/SPHERE}
   \titlerunning{Searching for \prim~f with Star-hopping RDI on SPHERE} 
   
   \author{
     Z. Wahhaj \inst{1,5} \and 
     J. Milli \inst{1,2} \and 
     C. Romero \inst{1,2} \and 
     L. Cieza \inst{3,4} \and 
     A. Zurlo \inst{3,4,5} \and 
     A. Vigan \inst{5} \and 
     E. Pe\~{n}a \inst{1} \and 
     G. Valdes \inst{1} \and 
     F. Cantalloube \inst{6} \and 
     J. Girard \inst{7} \and 
     B. Pantoja \inst{8}
   }
   \institute{
     European Southern Observatory, Alonso de C\'{o}rdova 3107,  Vitacura, Casilla 19001, Santiago, Chile\label{inst1} \\ \email{zwahhaj@eso.org} \and 
     Universit\'{e} Grenoble Alpes, CNRS, IPAG, F-38000 Grenoble, France\label{inst2} \and 
     N\'{u}cleo de Astronom\'{i}a, Facultad de Ingenier\'{i}a y Ciencias, Universidad Diego Portales, Av. Ejercito 441, Santiago, Chile\label{inst3} \and 
     Escuela de Ingenier\'{i}a Industrial, Facultad de Ingenier\'{i}a y Ciencias, Universidad Diego Portales, Av. Ejercito 441, Santiago, Chile\label{inst4} \and 
     Aix Marseille Univ, CNRS, CNES, LAM, Marseille, France\label{inst5} \and 
     Max Planck Institute for Astronomy, K\"{o}nigstuhl 17, 69117 Heidelberg, Germany\label{inst6} \and 
     Space Telescope Science Institute, 3700 San Martin Dr., Baltimore, MD 21218, USA\label{inst7} \and 
     Department of Physics and Astronomy, Bucknell University, Lewisburg, PA 17837\label{inst8}
   }
   \date{Received June 29, 2020; accepted December 20, 2020}

 
  \abstract
   {The direct imaging of extrasolar giant planets demands the highest possible contrasts ($\Delta$H~$\gtrsim$10~magnitudes) at the smallest angular separations ($\sim0.1''$)  from the star. We present an adaptive optics observing method, called {\it star-hopping}, recently offered as standard queue observing (service mode) for the SPHERE instrument at the VLT. The method uses reference difference imaging (RDI) but unlike earlier works, obtains images of a reference star  for PSF subtraction, within minutes of observing the target star.}
   {We aim to significantly gain in contrast over the conventional angular differencing imaging (ADI) method, to search for a fifth planet at separations less than 10~au, interior to the four giant planets of the HR~8799 system. The most likely semi-major axes allowed for this hypothetical planet, estimated by dynamical simulations in earlier work, were 7.5 and 9.7~au within a mass range of 1--8\mjup.}
   {We obtained 4.5 hours of simultaneous low resolution integral field spectroscopy (R$\sim$30, Y--H band with IFS) and dual-band imaging (K1 and K2-band with IRDIS) of the \prim\ system, interspersed with observations of a reference star. The reference star was observed for about one-third of the total time, and generally needs to be of similar brightness ($\Delta$R$\lesssim $1~magnitude) and separated on sky by $\lesssim$1--2$^o$. The hops between stars were made every 6--10 minutes, with only 1 minute gaps in on-sky integration per hop.} 
   {We did not detect the hypothetical fifth planet at the most plausible separations, 7.5 and 9.7 au, down to mass limits of 3.6 and 2.8\mjup\ respectively, but attained an unprecedented contrast limit of 11.2 magnitudes at 0.1$''$. We detected all four planets with high signal-to-noise ratios. The YJH spectra for planets $c$, $d$ were detected with redder H-band spectral slopes than found in earlier studies. As noted in previous works, the planet spectra are matched very closely by some red field dwarfs. Finally, comparing the current locations of the planets to orbital solutions, we found that planets $e$ and $c$ are most consistent with coplanar and resonant orbits. We also demonstrated that with star-hopping RDI, the contrast improvement at 0.1$''$ separation can be up to 2 magnitudes.}
   {Since ADI, meridian transit and the concomitant sky rotation are not needed, the time of observation can be chosen from within a 2–-3 times larger window. In general, star-hopping can be used for stars fainter than R=4 magnitudes, since for these a reference star of suitable brightness and separation is usually available.
     The reduction software used in this paper has been made available online$^1$.
   } 
         
   
   \maketitle

\section{Introduction}
\footnotetext[1]{\href{https://github.com/zwahhaj/starhopping}{https://github.com/zwahhaj/starhopping}.}

Radial velocity (RV) surveys have revealed to us the exoplanet population orbiting within $\sim$5~au of their parent stars \citep{2011arXiv1109.2497M, 2019ApJ...874...81F}. Transit techniques have done the same for the population of closer-in planets ($\lesssim$1~au), providing us a glimpse of their atmospheres as inferred from their spectra \citep{2010Sci...330..653H,2013ApJ...778...53D,2019ARA&A..57..617M}. Direct imaging on the other hand has found more than a dozen planets orbiting farther than 10~au from their stars (http://exoplanet.eu/). Direct imaging and interferometry are the only methods that allow us to obtain spectra of exoplanets separated by more than a few au from their host stars \citep{2014A&A...567L...9B, 2016A&A...587A..58B}. Direct imaging is also the only technique that captures protoplanetary disks in the act of forming planets \citep{2018A&A...617A..44K, 2018A&A...617L...2M, 2019NatAs...3..749H}. Moreover, it has shown us fully formed planetary systems with their left-over dusty planetesimal disks \citep{2012A&A...542A..40L}, and captured these dust-producing rocky disks at various stages over their lifetime \citep[e.g.,][]{2018A&A...614A..52B, 2020A&A...637L...5B,2016A&A...596L...4W}. 

Studies of systems like HR 8799 with its four planets can offer us a glimpse at possible early (age $<$ 30~Myrs) architectures \citep{2010Natur.468.1080M}, perhaps at a stage prior to major planet-migration and scattering \citep{2008ApJ...686..580C,2009sf2a.conf..313C, 2010ApJ...711..772R}. However, the extrasolar Jupiter and Saturn analogs are mostly still hidden from us, orbiting in the glare of their parent stars between 5 and 10~au \citep{2019ApJ...874...81F}.
Fortunately, a giant planet at an age of 30~Myr can be a hundred times brighter than at 300~Myr  \citep[e.g.,][]{2012RSPTA.370.2765A}.
With direct imaging, we are trying to detect the younger component of this hidden population, bridging the unexplored gap to connect to the RV and transit exoplanet populations closer in. In fact, some of the state-of-the-art direct imaging surveys have nearly completed and yielded a few more giant planets, fainter and orbiting closer to their stars than in earlier surveys, but mostly they report that the regions beyond 10~au rarely have planets more massive than 3--5\mjup. \citep{2019AJ....158...13N,2017A&A...605L...9C,2015Sci...350...64M}.

Especially for gound-based instruments, the success of the direct imaging technique, imaging dozens of exoplanets and protoplanetary disks has been mainly due to angular and spectral difference imaging \citep[ADI, SDI and ASDI;][]{2004Sci...305.1442L,2006ApJ...641..556M,2002ApJ...578..543S,2011ApJ...729..139W}. Without point spread function (PSF) differencing, within minutes we hit a wall in terms of sensitivity because of quasi static speckles in adaptive optics images. Speckles essentially mimic astronomical point sources, integrating more like signal than noise. The ADI, SDI and other related techniques decouple the speckles from the real signal, allowing them to be isolated and subtracted. However, these techniques are hampered by the self-subtraction problem \citep{2006ApJ...641..556M}. Since the decoupling of speckles and astronomical signal is never complete, there is inevitably some self-subtraction of signal. This can be manageable for planets moderately separated from the star, where we just lose sensitivity depending on the subtraction algorithm used \citep[e.g.,][]{2013ApJ...779...80W,2015A&A...581A..24W}. However, for planet-star separations of 1--2 resolution elements and extended structures like circumstellar disks, the signal can be completely subtracted or the morphology significantly altered or completely masked \citep{2012A&A...545A.111M}.

Reference difference imaging (RDI), a possible solution, has been routinely used in space telescope observations \citep[e\.g\.][]{1999ApJ...525L..53W, 2016ApJ...817L...2C}, as the PSF is quite stable over successive orbits of the telescope.  However, RDI is not often used in ground-based observing where PSFs change significantly over hours. This is because, prior to extreme AO, the PSF of other stars could not closely match the target PSFs in speckle similarity, especially if the reference star images were not obtained the same night as the science target. Nevertheless, impressive ground-based results on quite a few targets have been achieved \citep[][]{, 2009A&A...493L..21L, 2018AJ....156..156X, 2019AJ....157..118R,2020MNRAS.492..431B}. In the more recent efforts, reference PSFs were obtained 30 mins to hours apart and the telescope operator would have to manually setup the guiding for each target change, costing significant human effort and photon dead-time. Starting recently at VLT/SPHERE, we now offer fast automated RDI available in queue mode for the first time, requiring only a one minute gap for each target change, a technique monikered star-hopping RDI. To demonstrate the power of this new observing mode, and to look for new planets closer to the star, we targeted HR 8799, the home of the four giants.

HR~8799 is a young main-sequence star \citep[age 20--160~Myrs;][]{1969AJ.....74..375C,2006ApJ...644..525M, 2008Sci...322.1348M,2010ApJ...716..417H,2011ApJ...732...61Z,2012ApJ...761...57B} at a distance of 41.29$\pm$0.15~pc \citep{2018yCat.1345....0G}. The space motions of the star suggest membership in the Columbia moving group \cite[age 30--40~Myr;][]{2008hsf2.book..757T,2011ApJ...732...61Z,bell_mamajek_naylor_2015,2019MNRAS.483..332G}. It has four directly imaged giant planets at projected distances of 15, 27, 43, and 68~au \citep{2008Sci...322.1348M,2010Natur.468.1080M}. Upper-limit to the masses from orbital stability requirements and the derived luminosities assuming an age of $\sim$30~Myrs suggest that the planet masses are 5--7\mjup\ \citep{2010Natur.468.1080M,2011ApJ...729..128C,2012ApJ...755...38S}. Interior and exterior to the planets, warm dust at 6--10~au and an exo-Kuiper Belt beyond 100~au have been detected \citep{Sadakane_1986,Su_2009,2011ApJ...740...38H, 2014ApJ...780...97M,2016MNRAS.460L..10B}. Thus, it is likely that the planets formed in a circumstellar disk, instead of directly from a protostellar cloud as in binary or multiple star formation. However, it is currently a theoretical challenge to form so many massive planets in a single system. 

The total system architecture and stability, considering the age, mass and debris disk formation history have been studied in some detail  \citep[see][]{2009MNRAS.397L..16G,2014MNRAS.440.3140G,2018ApJS..238....6G,2009A&A...503..247R,Su_2009,2010ApJ...710.1408F,2010ApJ...717.1123M,2011ApJ...739L..41G,2014MNRAS.437.1378M,Matthews_2013,2016MNRAS.460L..10B,2016AJ....152...28K,2018ApJ...855...56W,2019MNRAS.483..332G}. \prim\ is a star of the $\lambda$~Bootis type (indicating an iron poor atmosphere), and also a $\gamma$~Dor variable, indicating small surface-pulsations perhaps also due to some accretion-associated chemical peculiarity \citep{2018MNRAS.477.2183S,2019MNRAS.487.2177S,2020A&A...635A.106T}. Spectra of the planets has been obtained in the NIR bands with Keck/OSIRIS \citep{2011ApJ...733...65B,2015ApJ...804...61B,2013Sci...339.1398K}, Project 1640 at Palomar \citep{2013ApJ...768...24O},   VLT/NACO \citep{2010ApJ...710L..35J}, GPI  \citep{2014ApJ...794L..15I} and SPHERE \citep{2016A&A...587A..57Z,2016A&A...587A..58B}. The comparison of the spectra to brown dwarfs, cool field objects and current atmospheric models suggest patchy thin and thick clouds of uncertain height, non-equilibrium chemistry, and a dusty low-gravity atmosphere \citep{2008Sci...322.1348M,2011ApJ...729..128C,2011ApJ...737...34M,2012ApJ...753...14S,2012ApJ...754..135M,2012ApJ...756..172M,2013ApJ...768..121A,2015ApJ...798..127B}. Given the theoretical challenge in explaining such a massive multi-planet and debris disk system with detailed and specific information, and the prospect of finding additional planets \citep{2014MNRAS.440.3140G,2018ApJS..238....6G} the system deserves a deeper look. We describe our SPHERE study of \prim\ in the following sections. The reduction software used in this paper can be found online \footnote{https://github.com/zwahhaj/starhopping}.

\section{Observations}
\subsection{Telescope and instrument control for Star-hopping}

The goal of star-hopping on VLT/SPHERE is to switch from recording adaptive optics corrected images of the science star to the reference star with only a $\sim$1 minute gap. Thus the usual help from the human operator to setup the guide star for the primary mirror's active optics correction, typically a 5 minute interaction, should be restricted to once per star, thus two times in total. This would allow us to make hops between science and reference star every $\sim$10 minutes without much loss in photon collecting efficiency, and ensuring minimal change in the PSF shape in the elapsed time. We do not provide an exact calculation for the optimum hopping frequency as it depends strongly on how the seeing and coherence time vary over the observation. However, we found in our observations that PSF similarity drops $\sim$2\% every 10 minutes (see Section~\ref{sec_rdi_vs_adi}). This is significant as the sensitivity reached depends non-linearly on the PSF subtraction quality. Thus, we recommend observing the science target for 10 minutes, then hopping to the reference star and observing it for 5 mins, repeating the cycle as needed.

To preserve PSF similarity and for time-efficiency, the AO loops would not be re-optimized when changing stars, and thus the reference star would need to have an R-magnitude (mag) within 1~mag of the science star, to ensure similar AO performance. While, we do not have strong constraints on the color of the reference star, again similar brightness (within 1~mag) in the observing wavelength is important. This is because the adaptive optics performance need to be similar and the signal-to-noise of the reference images need to be comparable or better. Also, the reference star would need to be within 1 to 2 degrees of the science star, so that the main mirror's shape-changes at the new pointing would not result in large changes in PSF properties. Fortunately, for the vast majority of stars fainter than $R\sim4$ mags a suitable reference star can be found, making star-hopping practical for $R\sim$~4--13~mag stars. The solution required new software to be designed for telescope control and new template software to be written for the observing sequence and instrument, i.e., SPHERE's control.  

For SPHERE, we designed two new acquisition templates called {\it starhop} and {\it hopback} which are only responsible for moving the telescope between the two stars and store relevant setup information so that subsequent hops can be made automatically. Thus a typical observing sequence would be: 1) Normal acquisition of science star with desired instrumental mode and setup, 2) An observing template lasting a few minutes, 3) Acquisition of reference star a few degrees away, with the {\it starhop} template, 4) Another observing template, 5) Quick return to the science star using the {\it hopback} template lasting $\sim$1 minute, 6) Another observing template 7) Quick return to reference star using the {\it hopback} template again, 8) As many iterations of steps 4 to 7 as desired. 

All three types of acquisitions constitute a full preset of the telescope, i.e., the primary mirror's shape and the secondary's pointing are set by a look-up table, then a guide star is selected (automatic for {\it hopback}) for accurate pointing corrections, continuous active optics corrections for the main mirror shape are activated using the guide star. However, human operators only assist with the first (normal) acquisition and the {\it starhop} acquisition, especially in the selection of the guide star and related setup. The {\it starhop} template stores all parameters required for these setups for the first star, moves (presets) to the second star, lets the operator assist in the second acquisition, and then stores all the parameters for the second acquisition. Small telescope offsets for fine-centering made by the operator when positioning the star on the instrument detector, are also recorded. Thus, the {\it hopback} template already has the relevant parameters saved and can automatically hop back and forth between the two stars, taking only $\sim$1 minute each time.

\subsection{HR~8799 observations}
We observed \prim~as part of a director's discretionary time (DDT) proposal, to test the performance limits of star-hopping with RDI on SPHERE. The SPHERE instrument \citep{2019A&A...631A.155B}, installed at the Nasmyth Focus of unit telescope 3 (UT3) at the VLT, is a state-of-the-art high-contrast imager, polarimeter and spectrograph, designed to find and characterize exoplanets. It employs an extreme adaptive optics system, SAXO \citep{2005OptL...30.1255F,2006OExpr..14.7515F,2012SPIE.8447E..1ZP,2016JATIS...2b5003S}, with 41$\times$41 actuators (1377 active in the pupil) for wavefront control, a low read noise EMCCD running at 1380 Hz, a fast (800 Hz bandwidth) tip-tilt mirror (ITTM) for pupil stabilization, extremely smooth toric mirrors \citep{2012A&A...538A.139H}, and a differential tip-tilt loop for accurate centering in the NIR. This system can deliver H-band strehl ratios for bright stars (R$<$9) of up to 90\% and continue to provide AO correction for stars as faint as R$=$14 mags. SPHERE also provides coronagraphs for diffraction suppression, including apodized Lyot coronagraphs \citep{2005ApJ...618L.161S} and achromatic four-quadrants phase masks \citep{2008A&A...482..939B}. It is comprised of three subsystems: the infrared dual-band imager and spectrograph \citep[IRDIS;][]{2008SPIE.7014E..3LD}, an integral field spectrograph \citep[IFS;][]{2008SPIE.7014E..3EC} and the Zimpol imaging polarimeter \citep[ZIMPOL;][]{2018A&A...619A...9S}.

We observed \prim~ in the IRDIFS extended mode \citep{2014A&A...572A..85Z}, where IRDIS K1 and K2-band images and IFS Y--H spectra are obtained simultaneously \citep{2010MNRAS.407...71V}.  The IRDIFS data was obtained in three 1.5 hour observing blocks (OBs), one block on the night of October 31, 2019 and two contiguous blocks on the night of November 1, 2019. We used the N\_ALC\_YJ\_S coronagraph with a central obscuration of radius 73~mas, which is not ideal for the maximum contrast in K-band but ensures that any object at 100~mas separation would not be partially obscured. With IRDIS we used 8s exposures, while with IFS we used 32s. We also obtained short-exposure unsaturated non-coronagraphic observations of the primary star for flux calibration, which we will call {\it FLUX} observations henceforth. The datasets can be found in the ESO archive under program~ID 2103.C-5076(A) and container IDs: 2622640, 2623891 and 2623923. Each container represents a separation epoch, consisting of several OBs alternating between \prim~ and the reference star. The reference star, HD~218381 (spectral type K0 vs F0V for HR~8799), is separated 0.55$^o$ from \prim~and is 0.52~mag fainter than it in R-band but 0.75~mag brighter in H-band. In total, we had 1440 IRDIS exposures for \prim~and 830 for the reference star. With IFS, we had 190 exposures for \prim~and 114 for the reference star. The observing conditions were average, with a coherence time of 4.7$\pm$1.3~ms, a seeing of 0.9$\pm$0.15$''$, and a windspeed of 2.1--7.7~m/s without the low-wind effect \citep{2018SPIE10703E..2AM}. The total sky-rotations were 23.8$^o$ on the first night and 53.4$^o$ on the second night. 

\section{Data Reduction}

\subsection{IFS reduction and contrast limit estimates}


Since our main motivation is to achieve sensitivities to fainter planets than earlier observations, we begin by estimating the detection limits of our data set and post-processing method. The detection limits are estimated by comparison to simulated planets which undergo the same reduction processes as the real planets. The measurement and analysis of the real planets in the system are presented afterwards.
For the basic reduction calibrations, we used SPHERE pipeline version 0.36.0 and scripts by \cite[][\it{http://astro.vigan.fr/tools.html}]{2015MNRAS.454..129V}
The IFS data sets from all 3 epochs were combined to form a cube of 7254 images, 186 images in each of the 39 wavelength channels. In each image, 16 simulated companions were inserted with offsets wrt. to the star, given by separations:  0.1$"$ to 1.6$"$ with steps of 0.1$"$ and position angles increments of 90$^o$ with each step.  The simulated companions were made from the {\it FLUX} exposures of the primary appropriately scaled in intensity. Since these sources were given constant chromatic contrast, i.e. the same spectra as the host star, we did not apply any spectral differencing in the reduction described below. The contrasts of these sources were chosen to be roughly 2 mags brighter than a preliminary contrast limit estimate for the data set. The reference PSF data set consisted of 4446 images. All science and reference images were unsharp-masked, i.e., each image was convolved with a Gaussian of FWHM 0.1$"$ (roughly twice the image resolution) and subtracted from the original to produce an image where most large scale spatial features like diffuse stellar light has been removed \citep[e\.g\.][]{1999PASP..111..587R,2013ApJ...779...80W}.

A diagonally oriented stripe pattern was found in all the IFS images, which we were unable to remove in the basic calibrated images. A zero-valued image passed through the basic calibration also yielded this pattern, found to be independent of the channel wavelength. Thus the pattern is likely an artefact of the pipeline. The output pattern image was bad-pixel cleaned and unsharp-masked to prepare it to be subtracted from the science images. Two annular regions were defined to optimize PSF subtraction, i.e., minimize the residual RMS in each region. These two annuli had inner and outer radii of 0.075$"$ and 0.67$"$, and 0.67$"$ and 1.33$"$ respectively. The science images were median-combined without de-rotation to reveal the background stripe pattern more clearly. Then we obtained the best intensity-scaled pattern images for the inner and outer annuli, which we in turn subtracted from each science image, to perform a preliminary removal of the pattern. Next, for each science image, we computed the best linear combination of reference images that reduced the RMS in the two annular regions separately, similar to the LOCI algorithm \citep{2007ApJ...660..770L}, but a much simpler version since optimization is done only over the two large annuli. We then took the difference of the science image and this composite reference image, and further applied an azimuthal profile removal filter as described in \citet{2013ApJ...779...80W}. All the difference images were median-combined again to check for any residual striped pattern, and remove it again by the same procedure as before.

Generally, we see a consistent but modest improvement in contrast ($\sim$ 0.2 mag) with the use of image filters (e.g. unsharp masking), and so we recommend their use. Also, we notice fewer artifacts, e.g. fewer PSF residuals in these reductions. However, as data sets may differ in PSF morphology, we also recommend studying reductions without applying such filters, even when trying to detect faint point sources.

   \begin{figure}[h]
   \centering
   \includegraphics[width=\hsize, trim = {0 0 0 5cm}, clip]{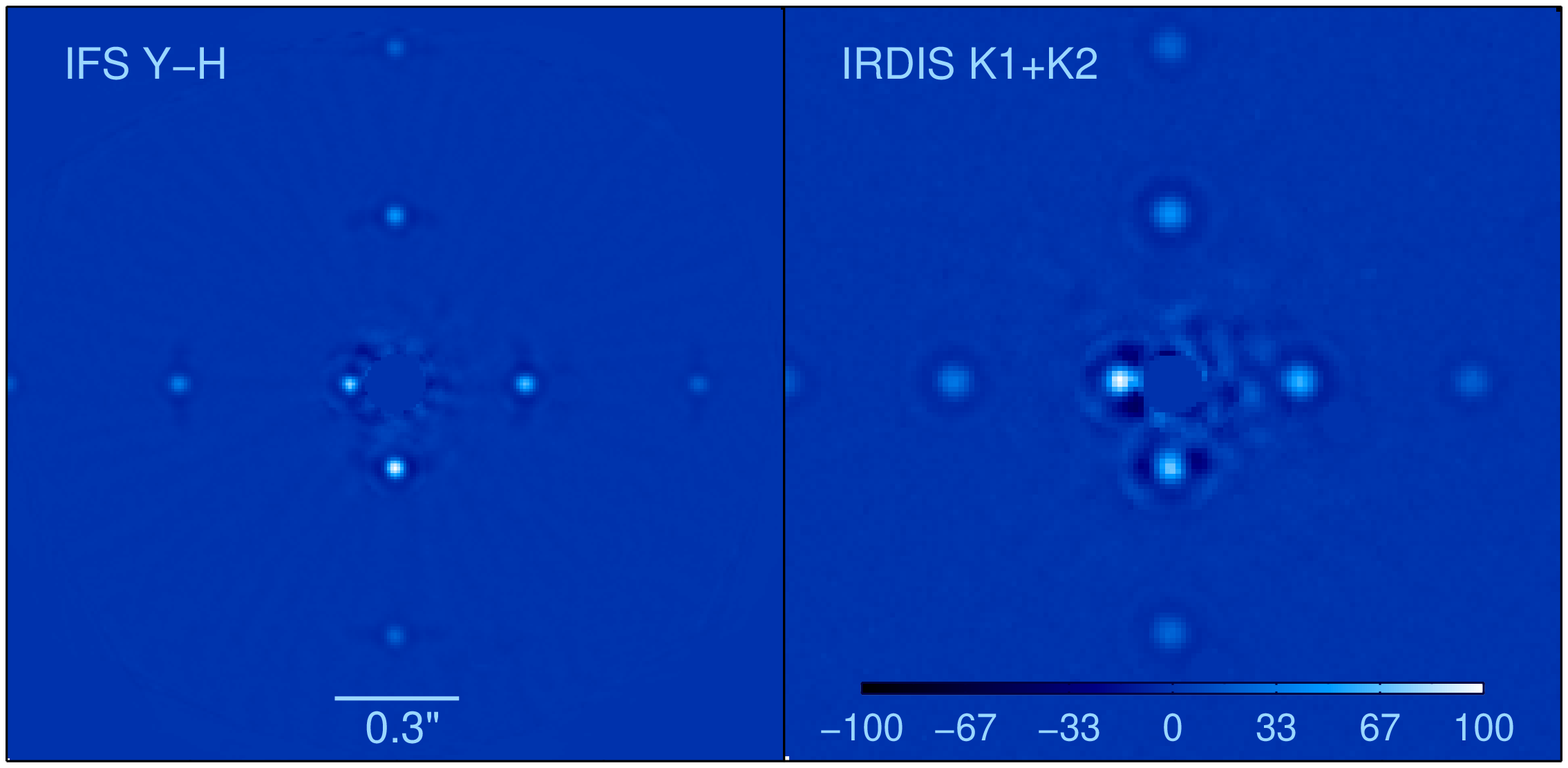}
      \caption{Left: An IFS Y--H band reduced image showing simulated planets which are recovered with high SNR. The source recovered closest to the star indicates a contrast limit of 11.2 mags at 0.1$''$ projected separation. Right: An IRDIS K1+K2 band reduced image also showing simulated planets at the same separations, all recovered with high SNR. The same contrast at 0.1$''$ was reached with IRDIS also. The planets were inserted into the basic calibrated data (flat-fielded, dark-subtracted and bad pixel corrected) All real planets have been masked out. The color scale is linear with intensity.
              }
         \label{fig_irdifs_fakes}
   \end{figure}

   \begin{figure}
   \centering
   \includegraphics[width=\hsize]{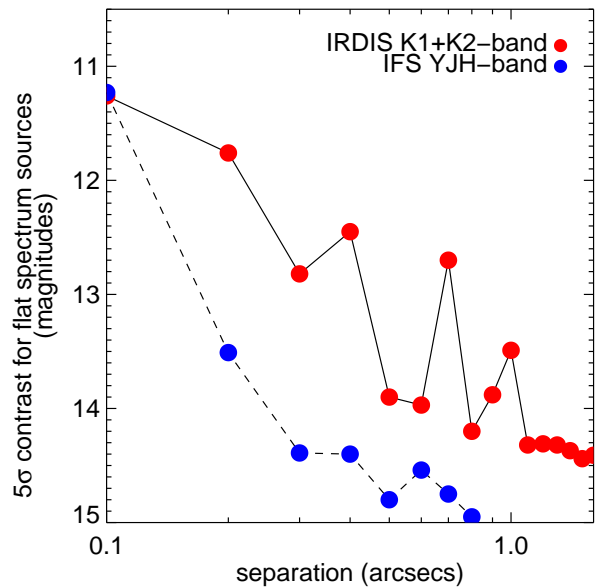}
      \caption{Contrast limits achieved in the IFS and IRDIS data sets, estimated by flux comparison to simulated planets recovered post-reduction.
              }
         \label{fig_contrast_irdifs}
   \end{figure}
   
\begin{figure*}[h]
   \includegraphics[width=\hsize, trim = {0 0 0 5cm}, clip]{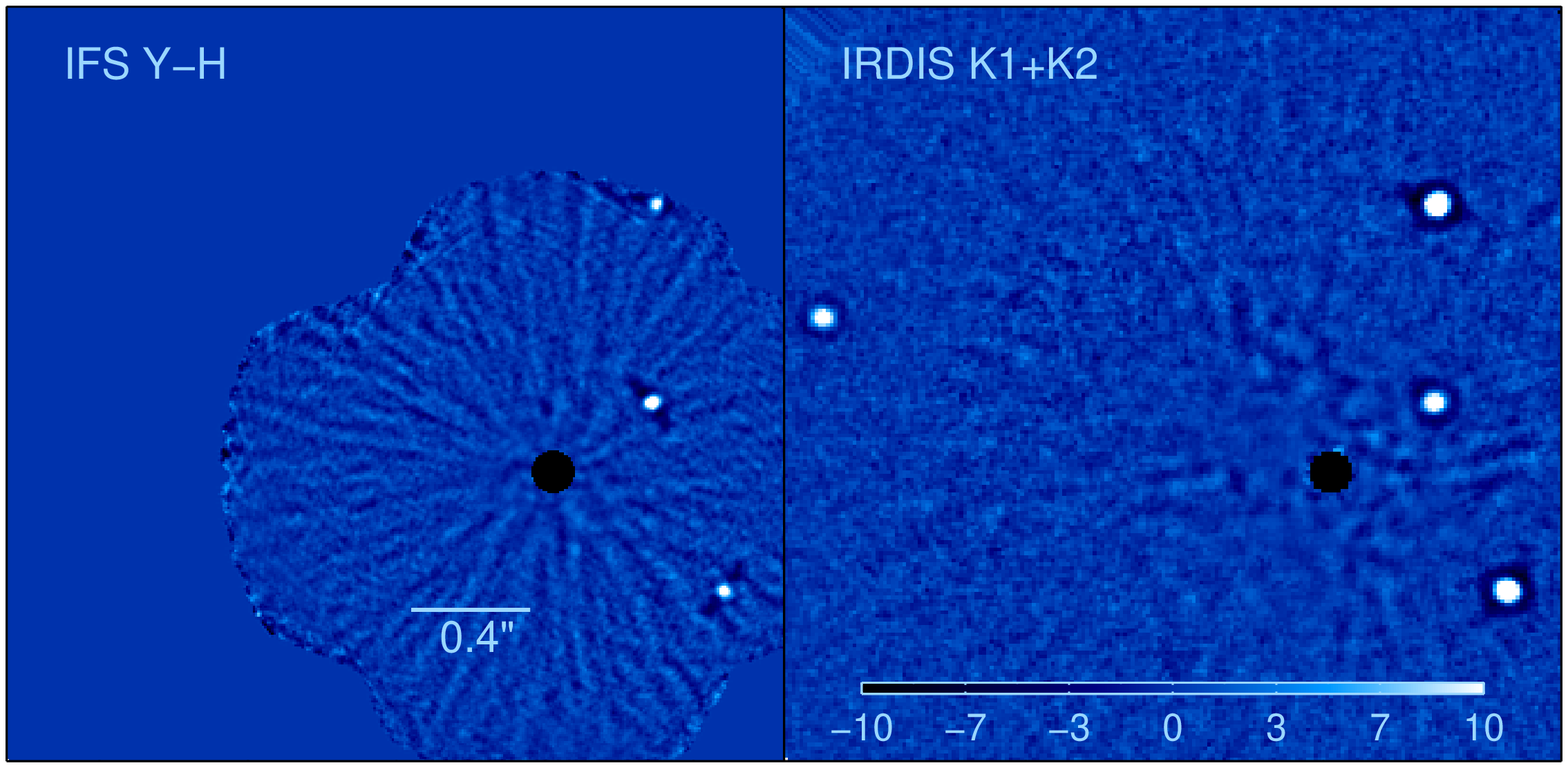}
      \caption{IFS and IRDIS images from star-hopping RDI reductions shown with same scale and orientation (North is up, East is left). Left: SNR map of the IFS Y--H band reduced image, showing only the real planets. The azimuthal filtering creates the dark negative arcs around the planets. They are more pronounced in the IFS reduction as more images were combined here than for the IRDIS reduction. Right: SNR map of the IRDIS K1+K2 band reduced image, showing only the real planets. The star, at the center of the black circle, is masked by the coronagraph. No new planets are detected in the newly probed region around 0.1$''$ separation above the contrast limit of 11.2 mags. }
         \label{fig_irdifs_nofakes}
   \end{figure*}
   
   Next, the images were derotated to align the sky with North up and East left orientation and median-combined. A signal-to-noise map is made for the final reduced image (Figure~\ref{fig_irdifs_fakes}), where the pixels in annular rings of width 4 pixels are divided by the robust standard deviation in that region. The robust value is taken to mitigate the effect of the simulated planets on the RMS. The signal-noise-ratio of each recovered simulated planet was then compared to its input contrast to calculate the 5$\sigma$-contrast limit achieved at the separation, like so $Contrast = InputContrast \times SNR/5$. The 5$\sigma$-contrasts achieved in this RDI-only reduction at 0.1$"$, 0.2$"$, 0.4$"$ and 0.8$"$ separations were 11.2, 13.5, 14.4 and 15 mags, corresponding to mass limits of 6.5, 3.1, 2.3 and 1.8\mjup\  respectively, as estimated from BT-Settl models assuming an age of 30~Myrs \citep{2012EAS....57....3A}. The contrast curve is shown in Figure~\ref{fig_contrast_irdifs}. The reduction showing only the real planets (without simulated planet insertions) is shown in Figure~\ref{fig_irdifs_nofakes}. No new planets are detected.


\subsection{IRDIS reduction and contrast limit estimates}
\label{sec_rdi_irdis}

The IRDIS reductions with simulated planets were done in a similar way to the IFS reductions. Since there were less images to process, we opted to use a more sophisticated but also more computation intensive reduction method. The simulated planets were inserted in the basic calibrated data at the same offsets with respect to the star as before. The planets inserted were $\sim$1 mag brighter than the 5$\sigma$ detection limit. For this exercise, we did not correct the relative rotational offset between IFS and IRDIS, so the PAs of the real HR~8799 planets do not agree between the two reduced images in Figure~\ref{fig_irdifs_fakes}. There were 1443 good science images in the three datasets combined and 828 reference images.

The images were first unsharp-masked. Next, we calculated the residual rms between all pairs of science and reference images, after intensity scaling to minimize the rms between 70~mas and 270~mas. For each science image, the best 16 reference images (more would worsen signal loss) were linearly combined by LOCI for subtraction to minimize the residual rms separately in annular rings covering the whole image.
Each target annulus, where the subtraction was actually done, had width 200mas. But the reference annuli, where LOCI tried to minimize the residual rms, started 25mas outside the target annuli and extended outwards to the cover the rest of the image. This was done to mitigate over-subtraction and signal loss. We chose these parameters mostly by trial and error. The azimuthal filtering, de-rotation and combination of all the difference images, and the contrast limit estimates were done in the same way as in the IFS reduction. The final reduced images (with and  without simulated companions) and the contrast performance are shown in Figures ~\ref{fig_irdifs_fakes},  ~\ref{fig_irdifs_nofakes} and ~\ref{fig_contrast_irdifs}, respectively. The IRDIS contrast limit is 11.2~mags at 0.1$''$ which is equal to the IFS limits, but IFS fares $\sim$0.5~mags better at larger separations.

\subsection{Comparison of RDI and ADI IRDIS detection limits}
\label{sec_rdi_vs_adi}
For a comparison of typical ADI and RDI IRDIS observations we use only the first of the three data sets, totalling 1.5 hours of execution time, since this is slightly longer that the typical observation length (1~hour) at the VLT. The data set constitutes 481 science images and for RDI, 276 reference images. The total sky rotation in the sciences images was 24$^o$. We performed 3 different ADI-based reductions which we call {\it ADI-LOCI-F1}, {\it ASDI-LOCI-F10} and {\it ASDI-PCA-F10}. The {\it ADI-LOCI-F1} is the same as the RDI reduction in terms of reference image selection and reference sector size and the use of LOCI, except that the references were restricted to those with more relative rotation than one-half FWHM (found by trial). The {\it ASDI-LOCI-F10} reduction (ASDI is Angular and Spectral Difference Imaging) was performed on a data set with simulated companions which were made 10 times fainter (thus labeled {\it F10}) in the K2 channel than in K1 channel, allowing aggressive spectral differencing and a potential contrast gain over ADI. Since reference images could have companions both spectrally and rotationally displaced, only the combined displacement need to be more than one-half FWHM. The {\it ASDI-PCA-F10} reduction was performed on the same data set as that of {\it ASDI-LOCI-F10}. The reduction parameters were again optimized by trial and error. We used principal component analysis (PCA) to construct the subtraction PSFs with 5 components \citep[See][]{2012ApJ...755L..28S}. However, for each science image, and for each annular sub-component of the image (same as the reductions above) only selected subsets of the science images were chosen as input for the PCA -- residual rms were calculated after subtracting all science image pairs, the best 30 matches (with least rms) that had more relative rotation than one-half FWHM were chosen, if less than 30 appropriate matches were found then the relative rotation criteria was relaxed to down to one-fourth FWHM, but no further, to allow input images for the PSF construction. This more selective approach to PCA helps to reduce the signal self-subtraction expected in ADI, and our tests supported this assumption, yielding significantly better results than PCA alone.

\begin{figure}[h]
   \centering
   \includegraphics[width=\hsize, trim = {0 0 0 5cm}, clip]{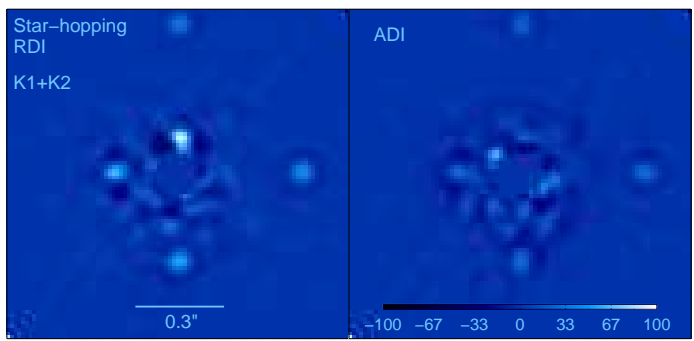}
      \caption{Comparison of star-hopping RDI versus ADI reductions of IRDIS K1+K2 band data injected with flat spectrum simulated planets. The inner two simulated planets are not successfully recovered in the ADI reduction, while they are clearly detected in the RDI reductions. The third simulated planet is recovered significantly better in the star-hopping RDI reduction. All real planets have been masked out. The color scale is linear with intensity.}
         \label{fig_contrast_irdis}
   \end{figure}

\begin{figure}
   \centering
   \includegraphics[width=\hsize]{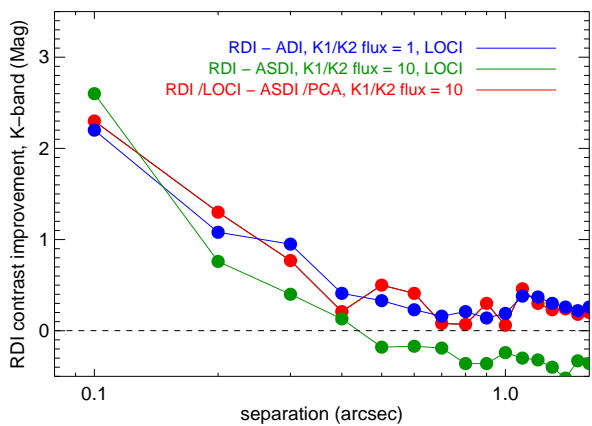}
      \caption{RDI contrast improvement over ADI or ASDI, estimated from the SNR of recovered simulated companions from an IRDIS data set. The star-hopping RDI technique yields detections limits more than 2 mags fainter than ADI at 0$''$.1 separation from the target star. The green line shows the case for a K1/K2 companion flux ratio of 10, and very similar algorithms for RDI and ASDI, except that the ASDI reduction is fine-tuned to minimize self-subtraction. The blue line similarly shows RDI$-$ADI difference for equal K1, K2 flux. The red line shows the RDI improvment against the best PCA-based ADI reduction for a K1/K2 flux ratio of 10. The LOCI and PCA reductions are described in section~\ref{sec_rdi_vs_adi}. }
         \label{fig_rdi_adi_contrast_irdis}
   \end{figure}

The RDI reduction (see top of section~\ref{sec_rdi_irdis}) was repeated for the same 1.5 hour data set used in the ADI reduction. In Figure~\ref{fig_contrast_irdis} we compare the RDI and the {\it ADI-LOCI-F1} reduction. The simulated planets inside 0.3$''$ separation are much better recovered in the star-hopping RDI reduction. In the ADI reduction, the innermost planet at 0.1$''$ is not recovered at all, while the one at 0.2$''$ is barely recovered. Contrast curves were calculated from the signal to noise ratio of the recovered simulated planets as before. The contrast improvement of RDI over the three ADI reductions, more than 2 mags at 0.1$''$ separation, is shown in Figure~\ref{fig_rdi_adi_contrast_irdis} as a difference between the two contrast curves. The improvement will of course vary with the total amount of sky rotation in the science images. 

Figure~\ref{fig_fracdiff_evol} illustrates why star-hopping RDI performs so much better than ADI. It shows the residual fractional rms (RFR) for each science image as a function of relative rotation, i.e., the remaining rms between 0.1$''$--0.3$''$ separations after subtraction of another science or reference star image, divided by the original rms in each science image. Specifically, $RFR_i=RMS(s_i-o_j)/RMS(s_i)$, where $s_i$ is a science image, $o_j$ is another science or reference star image and $RMS$ is computed between 0.1$''$--0.3$''$ separations. The RFRs post-RDI subtraction had a 2$\sigma$ range of 0.32--0.78. We see that although the science images provide better-matched PSFs in general, the images that can be used with minimal self-subtraction are much fewer and much poorer matches than the RDI reference set. Thus, the reference star images constitute a superior set for constructing subtraction PSFs.

\begin{figure}
   \centering
   \includegraphics[width=\hsize]{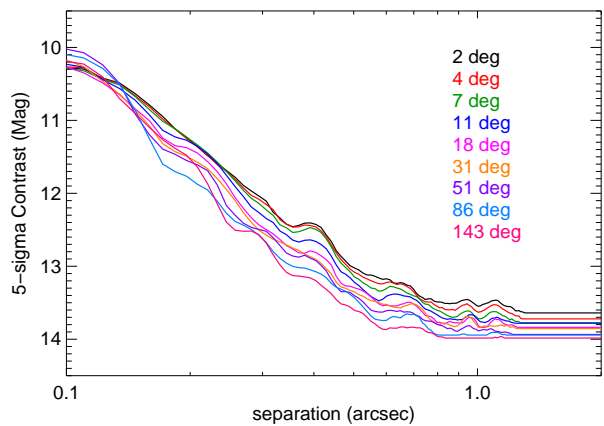}
      \caption{ Artificially increasing the field rotation for an RDI reduction before coadding the images does not improve the contrast significantly (see section~\ref{sec_rdi_vs_adi}). The legend gives the total rotation of the reduction for each contrast curve. At small separations (0.1--0.2$''$) we see no improvement, as contrasts are not correlated with rotation angles. At larger separations, we see a maximum of 0.5~mag improvement between the minimum and maximum rotations, 2$^o$ and 143$^o$, but only 0.3~mag improvement between 18$^o$ and 143$^o$.}
         \label{fig_rdi_rot_vs_con}
   \end{figure}

\begin{figure}
   \centering
   \includegraphics[width=\hsize]{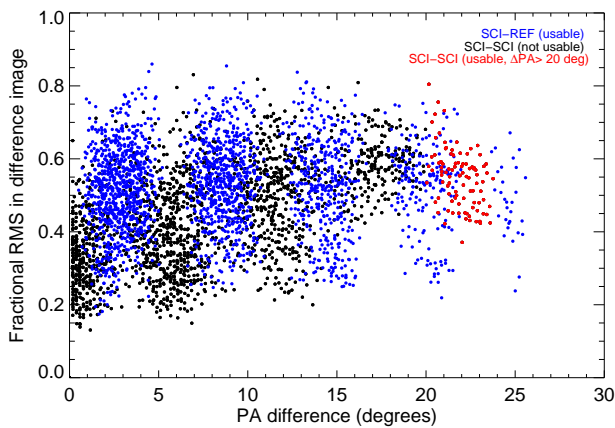}
      \caption{The comparison of PSF similarity between reference-science and science-science pairs. The residual fractional rms of difference images are plotted as a function of relative position angle/rotational offset. The black dots represent science-science subtractions, the blue dots represent science-reference subtractions, the red dots represent science-science differences with acceptable self-subtraction. For the science-reference points, the relevant quantity is the time difference, which in our case has an almost linear relationship to the PA difference.}
         \label{fig_fracdiff_evol}
   \end{figure}

In Figure~\ref{fig_rdi_rot_vs_con} we show that artificially increasing the field rotation for the RDI reduction (1.5 hour data set) before coadding the images does not improve the contrast significantly. Thus the speckle residuals are comparable to white noise as more rotation does not seem to result in additional smoothing. We estimate no improvement at 0.1$''$, $\sim$0.2 mag improvement farther out when comparing rotations of 140$^o$  to 20$^o$, and 0.5 mag improvement at 1$''$, when comparing rotations of 140$^o$  to 2$^o$. The reductions were done by mutliplying the actual position angles of the images by specific factors that would achieve total field rotaions of 2$^o$ to $\sim$140$^o$ (distributed logarithmically), before coadding the images.

During star-hopping tests on the night of August 8, 2019, we obtained 8 images for each of a pair of stars, HD~196963 and HD~196081, which are separated by $\sim$1.75$^o$. Since this pair has a much larger angular separation, we can use the RFR from this data set to gauge whether there is significant degradation in PSF similarity. Fortunately, the 2$\sigma$ range of the RFR was 0.33--0.53, indicating that star-hopping is still very effective for such large separations.
It should be noted that the coherence time was only 1.9--2.1~msec for these observations, compared to 2.5--7.2~msec for the HR~8799 observations. Although we have low statistics for such a performance, these results shows that even in poor to average conditions, star-hopping RDI can be effective for a pair of stars separated by almost 2$^o$.

\subsection{$JH$-band spectra from IFS}

The spectra of planets $c, d$ and $e$ were extracted with an aperture size of 3 pixels for all IFS channels. The spectra for planets $d$ and $e$ were corrected for flux loss by comparing them to three flat contrast sources (uniform contrast across wavelength) per planet inserted at the planets' separations, but at different PAs (offset from the planets by 30$^o$ to 270$^o$). These simulated planets are just the IFS $FLUX$ exposures scaled appropriately in intensity. They were inserted at 10 mags of contrast, which is somewhat brighter than the real planets. Since planet $c$ was detected at the edge of the IFS detector where simulated planets could not be inserted, we used the same comparison sources for planets $c$ and $d$. The simulated planets undergo the same reduction process as the real planets, and their fluxes are extracted using the same aperture sizes, and thus their systematic fractional flux error are the same. We verified this by checking that the spectrum recovered from the simulated companions did indeed have a uniform contrast. Thus the planet spectra is calculated as
\begin{equation}
B_{PR}(\lambda) = \frac{F_{PR}(\lambda)}{F_{PS}(\lambda)} \times 10^{-4} B_{S}(\lambda)
\end{equation}
where $F_{PR}$ and $F_{PS}$ are the real and simulated planet aperture fluxes respectively, and $B_S$ is the stellar spectra. Here, the fractional flux losses for the real planet are fully accounted for in the ratio, $F_{PR}(\lambda)/F_{PS}(\lambda)$.

\begin{figure}
   \centering
   \includegraphics[width=\hsize]{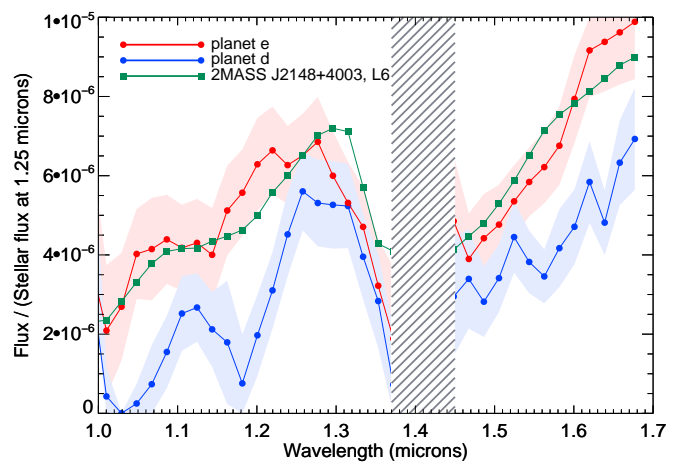}
      \caption{The spectra for planets $d$ and $e$ compared with that of the L6 object, 2MASS J2148+4003 from \citet{2008ApJ...686..528L}. The planet spectra have been divided by the stellar flux at 1.25~$\mu$m to show the contrast at that wavelength. The L6 object spectra was scaled to match planet $e$ at 1.25~$\mu$m. The shaded regions indicate the 1$\sigma$ error ranges of the spectra. The wavelength range 1.37--1.45~$\mu$m which is dominated by telluric lines is not shown.}
         \label{fig_ifs_spectra_de}
   \end{figure}

\begin{figure}
   \centering
   \includegraphics[width=\hsize]{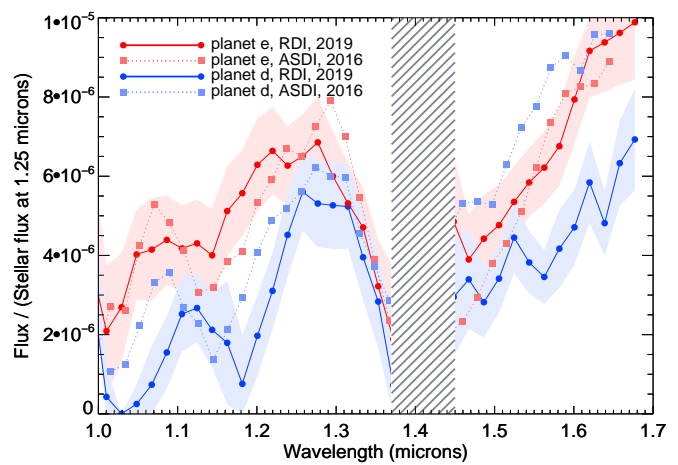}
      \caption{The RDI-extracted spectra for planets $d$ and $e$ in 2019 compared with their ADI-extracted spectra from 2016 as reported in \citet{2016A&A...587A..57Z}. The 2016 planet spectra to the 2019 have been matched at 1.25~$\mu$m for easier comparison for their respective shapes.  The shaded regions indicate the 1$\sigma$ error ranges of the spectra. The wavelength range 1.37--1.45~$\mu$m which is dominated by telluric lines is not shown.}
         \label{fig_ifs_spectra_de_2016comp}
   \end{figure}

\begin{figure}
   \centering
   \includegraphics[width=\hsize]{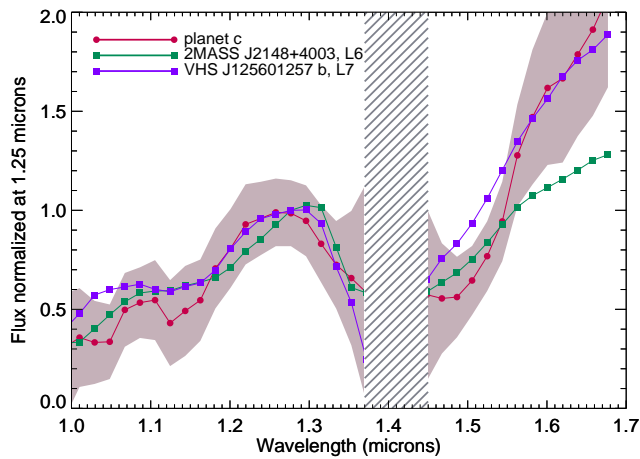}
      \caption{The RDI-extracted spectra for planets $c$  compared with that of the L6 object, 2MASS J2148+4003 from \citet{2008ApJ...686..528L} and L7 object VHS J125601257~b from \citet{2015ApJ...804...96G}. The wavelength range 1.37--1.45~$\mu$m which is dominated by telluric lines is not shown.}
         \label{fig_ifs_spectra_c}
   \end{figure}

The flux corrected spectra for planets $d$ and $e$ are shown in Figure~\ref{fig_ifs_spectra_de} along with that of the particularly red L6 object 2MASS J2148+4003 \citep{2008ApJ...686..528L} for comparison. All three spectra are much redder towards the $H$-band, in comparison to typical late L-types. Although not as red, the dusty dwarves of the field population also have redder than average spectra\citep[see ][]{2016A&A...587A..57Z, 2009ApJ...702..154S,2014ApJ...783..121G}. It should be noted that the spectra do differ somewhat in shape from earlier publications, \citep[e.g.,][]{2016A&A...587A..57Z}. This could be because the spectra we present here are the first not to be effected by signal self-subtraction due to ADI or SDI processing. The most notable differences from earlier spectra (see Figure~\ref{fig_ifs_spectra_de_2016comp}) are less defined peaks at 1.1$~\mu$m, and for planet $d$ in 2019, a gentler slope towards 1.6~$\mu$m. The absence of the peak at 1.1~$\mu$m is quite common among observed late L-type \citep[see Figure 3 of ][for example]{2016A&A...587A..58B}, and also seen in the spectra of 2MASS~J2148+4003. However, we also note that the 2016 spectral slopes towards 1.6~$\mu$m are very similar to planet $e$ in 2019. Although, the higher fluxes at 1.6~$\mu$m are rarer among such L-types, it would explain the earlier discrepancy between IRDIS and IFS fluxes near the $H$-band \citep{2016A&A...587A..57Z}. We could not estimate an accurate flux normalization for the spectra of planet $c$ as it was detected near the edge of the detector, so we show its spectra normalized to 1 at 1.25~$\mu$m in Figure~\ref{fig_ifs_spectra_c}. We do not pursue this further, as accurate $JH$-band photometry has already been provided in past publications. However, the shape of the planet's spectra is reliably detected and show's an even redder $J-H$ color than planets $d$ and $e$.  Although such red spectra are not common, a very similar slope (flux doubling between 1.25 and 1.6$\mu$m) was seen in the L7 object, VHS J125601257~b \citep{2015ApJ...804...96G}. This L7 object, a planetary candidate companion to a brown dwarf, is also thought to have a dusty atmosphere with thick clouds \citep[see ][ for a discussion]{2016A&A...587A..58B}.

\subsection{The \prim~debris disk}
\citet{2016MNRAS.460L..10B}, using the ALMA millimeter array, detected a broad debris ring, extending from $\sim$145~au to $\sim$430~au with an inclination of 40$\pm$6$^o$ and a position angle of 51$\pm$8$^o$. Prior to this, \citet{Su_2009} inferred from the spectral energy distribution of the system that a planetesimal belt extending from 100 and 300 au separation was the source of blow-out grains extending out to $\sim$1500~au. Thus the inner radius of the disk could start as close as 2.5$''$ and the outer radius could be as far as 11$''$ from the star.

\begin{figure}
  \centering
  \includegraphics[width=\hsize]{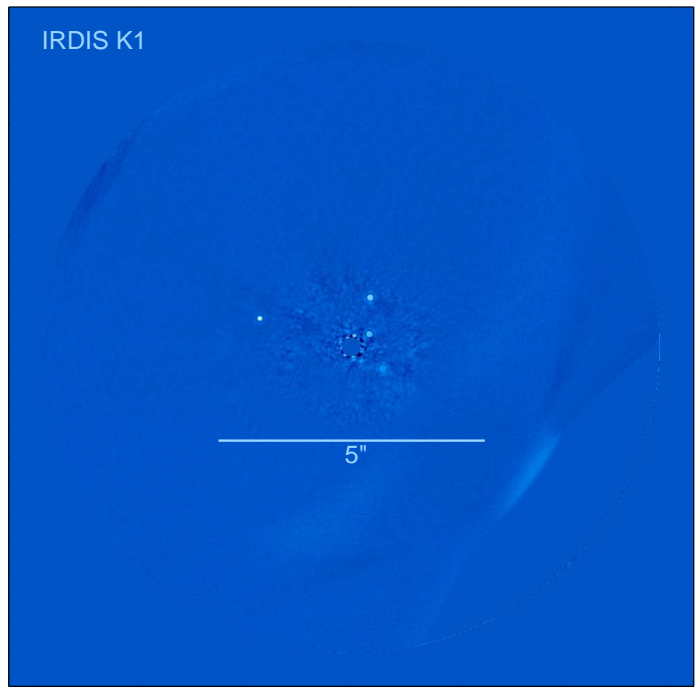}
  \caption{A IRDIS reduction without any prior image filtering to search for an extended circumstellar disk beyond angular separations of 2.5$''$ (to $>$6$''$) from the star. ALMA observations by \citet{2016MNRAS.460L..10B} indicate that the disk should have a position angle of 51$\pm$8$^o$ and an inclination of 40$\pm$6$^o$. We do not detect any disk down to a contrast limit of 14.1 magnitudes. Some faint thermal emission from the detector background is seen in the lower right, but not in the expected orientation of the known disk. North is up and East is to the left.} 
  \label{fig_irdis_disk}
\end{figure}

It is expected that RDI reductions would be a major improvement over ADI for detections of disks with large angular extents, as self-subtraction in these cases is a severe problem for ADI. To detect the disk, we repeated the IRDIS RDI reduction without simulated companions or any prior image filtering (used to enhance speckle subtraction), as these remove all extended emission. We only used the K1-band images as the K2-band have much higher background. Detecting disks which are close to azimuthally symmetric in the plane of the sky, and extended over several arcseconds is a challenge very different from planet recovery, as the expected signal area is most of the image and the background area is perhaps non-existent. The image sectors used for PSF subtraction cannot be small, as this would remove extended signal. So, we used one large annulus extending from 0.4$''$ to 2$''$ separations to cover most of the PSF halo. The final reduction is shown in Figure~\ref{fig_irdis_disk}, but no disk emission was detected down to a 5$\sigma$ contrast of 14.1 magnitudes beyond 2.5$''$ separations. The non-detection is not surprising given the marginal detection of the much brighter 49~Cet debris disk with SPHERE \citep{2017ApJ...834L..12C}. The fractional disk luminosity of \prim~is 8$\times$10$^{-5}$ \citep{2009ApJ...705..314S} versus 9$\times$10$^{-4}$ for 49~Cet \citep{2015MNRAS.447..577M}. The inner radius of the disks start at roughly 2$''$ separation for both \citep{2017ApJ...834L..12C, 2016MNRAS.460L..10B}, with expected physical separations of 100--150~au. The two stars have similar spectral types (F0--A1) with very similar $H$-band magnitudes (5.3--5.5~mag).

\subsection{IRDIS K1, K2 band photometry}
The photometry of the four planets were extracted by comparison with simulated planets in a similar way to the IFS spectra. For each of the four planets, three simulated planets were inserted into the dataset with a contrast of 10 mags, at the same separation as the real planets, but with large PA offsets (30 to 270$^o$). The relative aperture photometry was done similar to IFS, but with aperture radius 4 pixels, because of the larger FWHM in the K-band. The recovered photometry are all brighter than the \citet{2016A&A...587A..57Z} measurements by about 0.1 mag (see Table~\ref{table_astrom}). The standard deviation in the contrasts estimates for the three reference simulated planets is less than 0.03 mags. The dominant contrast uncertainty comes from the measurement of the AO-corrected stellar PSF core flux, which is measured only once every 1.5 hours.

\subsection{Astrometric measurements and comparison to orbital models}

The IRDIS data set for science images were separately reduced by the {\bf SPHERE data center} \citep{2017sf2a.conf..347D} which treated it as an ordinary pupil-tracking sequence. The data center applied the optimal distortion correction methods consistent with \citet{2016A&A...587A..56M}, to produce a basic-calibrated data set with high astrometric fidelity (3--4~mas). These images were then reduced using the high-contrast imaging algorithm, ANDROMEDA \citep{2015A&A...582A..89C}, to produce astrometric measurements (see Table~\ref{table_astrom}) for the four known HR~8799 planets. We also compared the recovered coordinates for the real planets between the RDI and ADI reductions, and found that the planet locations agreed to within 2.7~mas, smaller than the errors estimated in Table~\ref{table_astrom}.

\begin{figure}
  \centering 
  \includegraphics[width=\hsize]{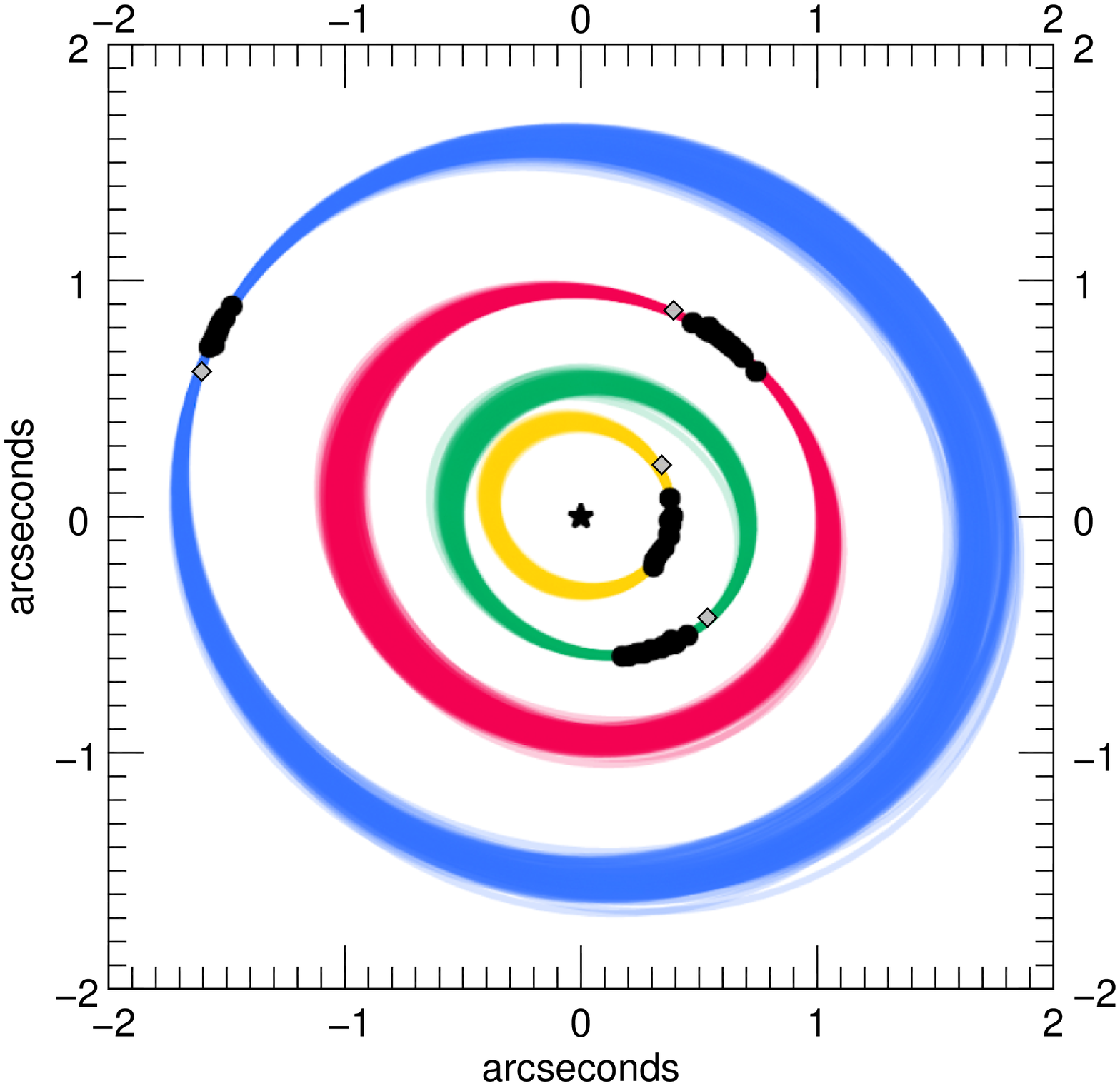}
  \hspace*{1cm}\includegraphics[angle=-90,width=9cm, trim = {0 5cm 0 0}, clip]{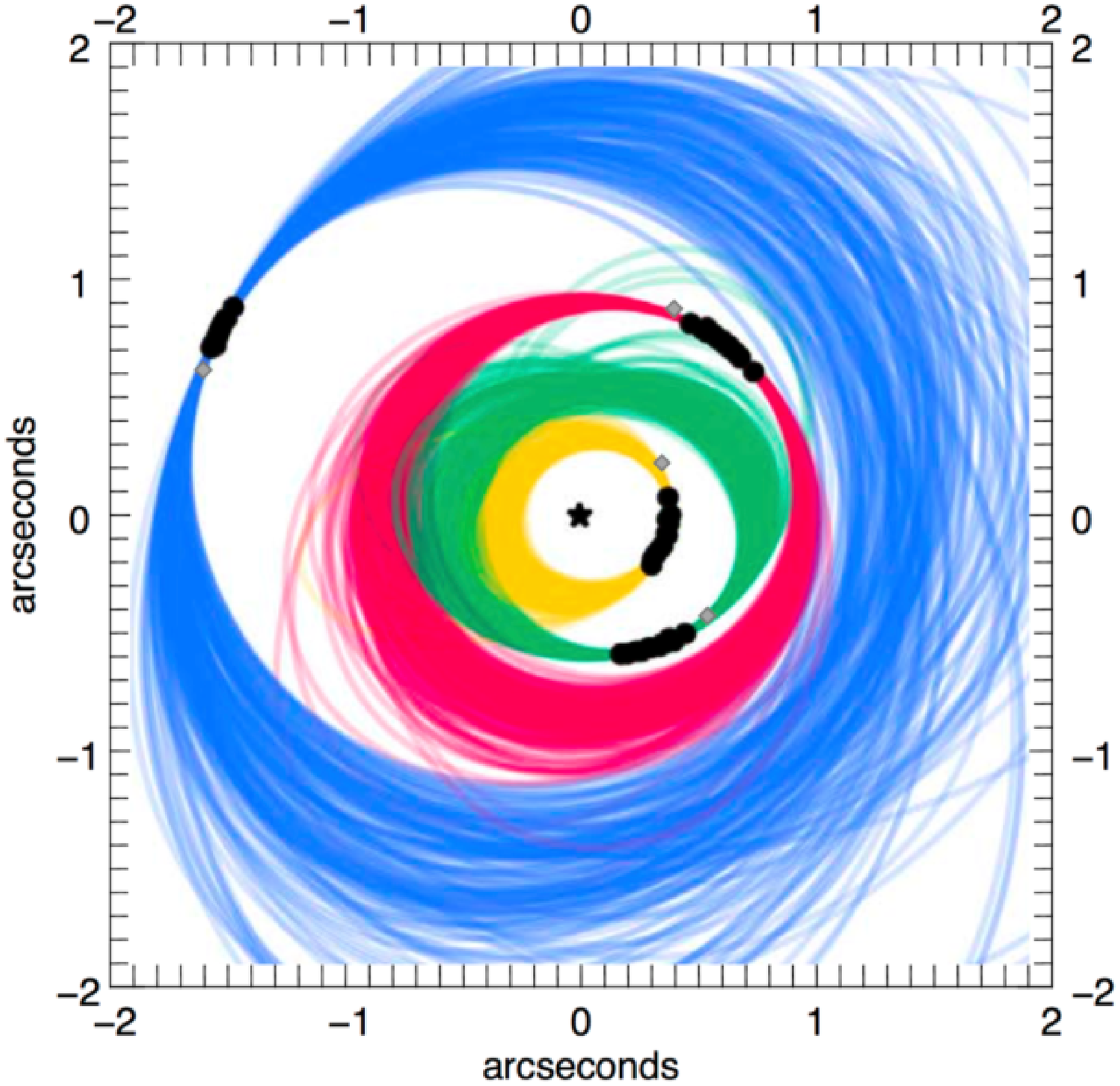}
  \caption{Top: The November 1, 2019 epoch astrometry overlaid as gray diamonds on the most dynamically stable orbital solutions from \citet{2018AJ....156..192W} (see their Figure 4), where coplanarity and 1:2:4:8 period commensurabilities were imposed. The black dots represent earlier measured astrometry for the four planets. Bottom: Same points overlaid on the orbital solutions without the additional constraints. The 2019 locations for planets $e$ and $c$ are more consistent with the dynamical stable family of orbits.}
  \label{fig_planet_orbits}
\end{figure}

An exhaustive orbital fitting effort is being currently undertaken by Zurlo et al.\ (in preparation) including all extant astrometry. Moreover, extensive work has been done to find orbital solutions to the prior astrometry for this system, so we just compare our latest measurements to the viable orbits computed by \citet{2018AJ....156..192W}. From millions of orbits generated by a monte carlo method, they generated 3 sets of solutions: 1) the orbits are forced to be coplanar and have 1:2:4:8 orbital commensurabilities, 2) no coplanarity but with low eccentricity and period commensurabilities as before, 3) with no additional constraints. In Figure~\ref{fig_planet_orbits}, we overlay our astrometry on orbital solution sets 1 and 3. Although, the latest points are consistent with both sets of solutions, planets $e$ and $c$ fall close to the expected position in the dynamically stable set, but a bit far from the mean expected location in the unconstrained set of orbits. Thus, the coplanar orbits with period commensurabilities are favored in our comparisons.

Survival of the four planets and even a hypothetical fifth planet is possible for the lifetime of the system ($>$ 30 Myrs), but requires the period commensurabilities mentioned above. In fact, this was needed even when only planets $b$, $c$ and $d$ were known \citep{2009MNRAS.397L..16G, 2009A&A...503..247R,2010ApJ...710.1408F, 2010IJAsB...9..259M}. Such dynamical models envision that the four planets were formed at larger separations and migrated inwards. This would allow the very similar chemical compositions indicated by their spectra, as opposed to more variation expected if they had formed insitu \citep{2010Natur.468.1080M}. 

The most likely semi-major axes allowed for the hypothetical inner planet $f$, estimated by \citet{2014MNRAS.440.3140G, 2018ApJS..238....6G} were 7.5~au and 9.7~au, with dynamical constraints on the masses of 2--8\mjup\ and 1.5--5\mjup\  respectively. The IFS contrasts we achieved at these separation were 13.05 and 13.86 mags, corresponding to estimated masses of 3.6\mjup\ and 2.8\mjup\ respectively (assuming an age of 30~Myr), from the BT-Settl models \citep{2012EAS....57....3A}. Thus, the planet may still exist with a mass of 2--3.6\mjup\ at 7.5~au or 1.5--2.8\mjup\ at 10~au. 

\begin{table*}
\caption{Astrometry and photometry of the four HR~8799 planets.}             
\label{table_astrom}      
\centering                          
\begin{tabular}{c c c c c c c c c}        
\hline\hline                 
planet   &  $\rho$ (mas) & $\sigma_{\rho}$(mas)  & PA      &     $\sigma_{PA}$ &  SNR   &     $\Delta$K1 (mag) &     $\Delta$K2 (mag) & Mass ($M_J$)\\
\hline
   e  &  406    & 4  & 302.72$^o$  &  0.04$^o$ & 41 &  10.8$\pm$0.02 &  10.63$\pm$0.03  & 8$^{+7}_{-2}$ \\
   d  &  686    & 4  & 231.38$^o$  & 0.006$^o$ & 83 &  10.7$\pm$0.02 &  10.47$\pm$0.02  & 8$^{+7}_{-2}$ \\
   c  &  958    & 3  & 335.86$^o$  &  0.05$^o$ & 96 &  10.8$\pm$0.02 &  10.53$\pm$0.03  & 8$^{+7}_{-2}$ \\
   b  & 1721    & 4  &  69.05$^o$  &  0.04$^o$ & 47 &  11.89$\pm$0.01 &  10.75$\pm$0.01 & 6$^{+7}_{-1}$ \\
\hline                                   
\end{tabular}
\\[10pt]
\footnotesize{The mass estimates are from the PHOENIX BT-Settl atmospheric models \citep{2015A&A...577A..42B}, assuming an age of 30$^{+130}_{-10}$ Myrs. However, the most dynamically stable orbital solutions from \citet{2018AJ....156..192W} set much tighter limits: a mass of $5.8\pm0.5~M_J$ for planet $b$, and $7.2\pm0.6~M_J$ for the other planets.}
\end{table*}

\section{Conclusions}

In this paper, we successfully used the new star-hopping RDI technique to detect all four known planets of the HR~8799 system, and significantly improved on the contrast limits attained previously with ADI, at separations less than 0.4$''$. This technique of moving quickly to a reference star to capture a similar AO PSF for differencing, with only a 1~minute gap in photon collection, can now be used in service mode at the VLT with all the observing modes available on the SPHERE instrument. Using star-hopping RDI, we demonstrated the contrast improvement at 0.1$''$ separation can be up to 2 mags, while at larger separations the improvement can be $\approx$1--0.5~mags, results which are comparable to those of \cite{2019AJ....157..118R}. With this technique there is no need for any local sidereal time constraints during observations, which is usually essential for ADI observations. This means that the observing window for typical targets can be expanded by a factor of 2--3. Moreover, star-hopping can usually be used for stars fainter than R$=$4~mag, as for these a reference star of comparable brightness can be found within 1--2~degrees (closer is better). Indeed we found comparable PSF similarity for a pair of stars 1.75$^o$ apart. The technique provides significant contrast improvement mainly because of two reasons: the usable PSF, those without significant self-subtraction or flux loss from PSF subtraction 1) occur closer in time and thus are more similar to the target image than in ADI and 2) are more numerous than in ADI as they are spread uniformly over the whole sequence, rather than only available after significant sky rotation. The benefit for extended object like disks will be the most impactful, as in ADI the self-subtraction artefacts can result in significant change in their apparent morphology.

In our SPHERE observations of HR~8799, we did not detect planet $f$ at the most plausible locations, 7.5 and 9.7~au, down to mass limits of 3.6 and 2.8\mjup, respectively. Also, we did not detect any new candidate companions, even at the smallest observable separation, 0.1$''$ or $\approx$~4.1~au, where we attained a contrast limit of 11.2~mags or 6\mjup\ in K1+K2-band \citep[6.5\mjup\ in JHK-band using BT-Settl models from][]{2012RSPTA.370.2765A}. However, we detected all 4 planets in K1+K2-band with SNR of 41, 83, 96 and 47 for planets $e$, $d$, $c$ and $b$, respectively. The YJH spectra for planets $c$, $d$, $e$ were detected with very red colors. Our spectra of  planet $c$ has higher SNR than earlier observations \citep[P1640,][]{2013ApJ...768...24O,2015ApJ...803...31P}. Planets $c$ and $d$ spectra have some differences with respect to earlier observations. Particularly, the spectral slope is redder in the H-band, which is significant as that part of the spectra has the highest SNR. This could be due to real evolution of the atmosphere of the planets over the past few years. Previous work has already shown that the spectra are difficult to find close matches with current compositional models due to inadequate understanding of cloud properties and non-equilibrium chemistry \citep{2016A&A...587A..58B}. However, the spectra are matched very closely by some red field dwarfs and a planetary mass companion to a brown dwarf \citep[VHS J125601257~b;][]{2015ApJ...804...96G}. We disk not detect the debris disk seen by ALMA \citep{2016MNRAS.460L..10B}, but this is not surprising given that the much brighter debris disk of a comparable system, 49~Cet, was only marginally detected by SPHERE \citep{2017ApJ...834L..12C}. Finally, comparing the current locations of the planets to orbital solutions from \citet{2018AJ....156..192W}, we found that planets $e$ and $c$ are more consistent with coplanar and resonant orbits than without such restrictions.  

In summary, the star-hopping RDI technique significantly boosts SPHERE's detection capabilities both for planets and circumstellar disks, and should contribute to high-impact exoplanet science, as the technique is brought to other telescope facilities.

\begin{acknowledgements}
  This work has made use of the the SPHERE Data Centre, jointly operated by OSUG/IPAG (Grenoble), PYTHEAS/LAM/CESAM (Marseille), OCA/Lagrange (Nice), Observatoire de Paris/LESIA (Paris), and Observatoire de Lyon. We would especially like to thank Nadege Meunier at the SPHERE data center in Grenoble for the distortion-corrected reductions used for the astrometric measurements, Bartosz Gauza for providing the spectra of VHS~J125601.92-125723.9~b, Jason Wang for allowing us to use the dynamical modeling figures from his publication, and Matias Jones, Florian Rodler, Benjamin Courtney-Barrer, Francisco Caceres and Alain Smette at the VLT for technical help during the various phases of the development of the star-hopping technique. 
 
\end{acknowledgements}

%
%

\bibliographystyle{aa}
\bibliography{zrefs}

\begin{thebibliography}{112}
\expandafter\ifx\csname natexlab\endcsname\relax\def\natexlab#1{#1}\fi

\bibitem[{{Allard} {et~al.}(2012{\natexlab{a}}){Allard}, {Homeier}, \&
  {Freytag}}]{2012RSPTA.370.2765A}
{Allard}, F., {Homeier}, D., \& {Freytag}, B. 2012{\natexlab{a}}, Philosophical
  Transactions of the Royal Society of London Series A, 370, 2765

\bibitem[{{Allard} {et~al.}(2012{\natexlab{b}}){Allard}, {Homeier}, {Freytag},
  \& {Sharp}}]{2012EAS....57....3A}
{Allard}, F., {Homeier}, D., {Freytag}, B., \& {Sharp}, C.~M.
  2012{\natexlab{b}}, in EAS Publications Series, Vol.~57, EAS Publications
  Series, ed. C.~{Reyl{\'e}}, C.~{Charbonnel}, \& M.~{Schultheis}, 3--43

\bibitem[{{Apai} {et~al.}(2013){Apai}, {Radigan}, {Buenzli}, {Burrows}, {Reid},
  \& {Jayawardhana}}]{2013ApJ...768..121A}
{Apai}, D., {Radigan}, J., {Buenzli}, E., {et~al.} 2013, \apj, 768, 121

\bibitem[{{Baines} {et~al.}(2012){Baines}, {White}, {Huber}, {Jones},
  {Boyajian}, {McAlister}, {ten Brummelaar}, {Turner}, {Sturmann}, {Sturmann},
  {Goldfinger}, {Farrington}, {Riedel}, {Ireland }, {von Braun}, \&
  {Ridgway}}]{2012ApJ...761...57B}
{Baines}, E.~K., {White}, R.~J., {Huber}, D., {et~al.} 2012, \apj, 761, 57

\bibitem[{{Baraffe} {et~al.}(2015){Baraffe}, {Homeier}, {Allard}, \&
  {Chabrier}}]{2015A&A...577A..42B}
{Baraffe}, I., {Homeier}, D., {Allard}, F., \& {Chabrier}, G. 2015, \aap, 577,
  A42

\bibitem[{{Barman} {et~al.}(2015){Barman}, {Konopacky}, {Macintosh}, \&
  {Marois}}]{2015ApJ...804...61B}
{Barman}, T.~S., {Konopacky}, Q.~M., {Macintosh}, B., \& {Marois}, C. 2015,
  \apj, 804, 61

\bibitem[{{Barman} {et~al.}(2011){Barman}, {Macintosh}, {Konopacky}, \&
  {Marois}}]{2011ApJ...733...65B}
{Barman}, T.~S., {Macintosh}, B., {Konopacky}, Q.~M., \& {Marois}, C. 2011,
  \apj, 733, 65

\bibitem[{Bell {et~al.}(2015)Bell, Mamajek, \&
  Naylor}]{bell_mamajek_naylor_2015}
Bell, C. P.~M., Mamajek, E.~E., \& Naylor, T. 2015, Proceedings of the
  International Astronomical Union, 10, 41–48

\bibitem[{{Beuzit} {et~al.}(2019){Beuzit}, {Vigan}, {Mouillet}, {Dohlen},
  {Gratton}, {Boccaletti}, {Sauvage}, {Schmid}, {Langlois}, {Petit},
  {Baruffolo}, {Feldt}, {Milli}, {Wahhaj}, {Abe}, {Anselmi}, {Antichi},
  {Barette}, {Baudrand}, {Baudoz}, {Bazzon}, {Bernardi}, {Blanchard}, {Brast},
  {Bruno}, {Buey}, {Carbillet}, {Carle}, {Cascone}, {Chapron}, {Charton},
  {Chauvin}, {Claudi}, {Costille}, {De Caprio}, {de Boer}, {Delboulb{\'e}},
  {Desidera}, {Dominik}, {Downing}, {Dupuis}, {Fabron}, {Fantinel}, {Farisato},
  {Feautrier}, {Fedrigo}, {Fusco}, {Gigan}, {Ginski}, {Girard}, {Giro},
  {Gisler}, {Gluck}, {Gry}, {Henning}, {Hubin}, {Hugot}, {Incorvaia}, {Jaquet},
  {Kasper}, {Lagadec}, {Lagrange}, {Le Coroller}, {Le Mignant}, {Le Ruyet},
  {Lessio}, {Lizon}, {Llored}, {Lundin}, {Madec}, {Magnard}, {Marteaud},
  {Martinez}, {Maurel}, {M{\'e}nard}, {Mesa}, {M{\"o}ller-Nilsson}, {Moulin},
  {Moutou}, {Orign{\'e}}, {Parisot}, {Pavlov}, {Perret}, {Pragt}, {Puget},
  {Rabou}, {Ramos}, {Reess}, {Rigal}, {Rochat}, {Roelfsema}, {Rousset}, {Roux},
  {Saisse}, {Salasnich}, {Santambrogio}, {Scuderi}, {Segransan}, {Sevin},
  {Siebenmorgen}, {Soenke}, {Stadler}, {Suarez}, {Tiph{\`e}ne}, {Turatto},
  {Udry}, {Vakili}, {Waters}, {Weber}, {Wildi}, {Zins}, \&
  {Zurlo}}]{2019A&A...631A.155B}
{Beuzit}, J.~L., {Vigan}, A., {Mouillet}, D., {et~al.} 2019, \aap, 631, A155

\bibitem[{{Boccaletti} {et~al.}(2008){Boccaletti}, {Chauvin}, {Baudoz}, \&
  {Beuzit}}]{2008A&A...482..939B}
{Boccaletti}, A., {Chauvin}, G., {Baudoz}, P., \& {Beuzit}, J.~L. 2008, \aap,
  482, 939

\bibitem[{{Boccaletti} {et~al.}(2020){Boccaletti}, {Di Folco}, {Pantin},
  {Dutrey}, {Guilloteau}, {Tang}, {Pi{\'e}tu}, {Habart}, {Milli}, {Beck}, \&
  {Maire}}]{2020A&A...637L...5B}
{Boccaletti}, A., {Di Folco}, E., {Pantin}, E., {et~al.} 2020, \aap, 637, L5

\bibitem[{{Boccaletti} {et~al.}(2018){Boccaletti}, {Sezestre}, {Lagrange},
  {Th{\'e}bault}, {Gratton}, {Langlois}, {Thalmann}, {Janson}, {Delorme},
  {Augereau}, {Schneider}, {Milli}, {Grady}, {Debes}, {Kral}, {Olofsson},
  {Carson}, {Maire}, {Henning}, {Wisniewski}, {Schlieder}, {Dominik},
  {Desidera}, {Ginski}, {Hines}, {M{\'e}nard}, {Mouillet}, {Pawellek}, {Vigan},
  {Lagadec}, {Avenhaus}, {Beuzit}, {Biller}, {Bonavita}, {Bonnefoy},
  {Brandner}, {Cantalloube}, {Chauvin}, {Cheetham}, {Cudel}, {Gry}, {Daemgen},
  {Feldt}, {Galicher}, {Girard}, {Hagelberg}, {Janin-Potiron}, {Kasper}, {Le
  Coroller}, {Mesa}, {Peretti}, {Perrot}, {Samland}, {Sissa}, {Wildi}, {Zurlo},
  {Rochat}, {Stadler}, {Gluck}, {Orign{\'e}}, {Llored}, {Baudoz}, {Rousset},
  {Martinez}, \& {Rigal}}]{2018A&A...614A..52B}
{Boccaletti}, A., {Sezestre}, E., {Lagrange}, A.~M., {et~al.} 2018, \aap, 614,
  A52

\bibitem[{{Bohn} {et~al.}(2020){Bohn}, {Kenworthy}, {Ginski}, {Manara},
  {Pecaut}, {de Boer}, {Keller}, {Mamajek}, {Meshkat}, {Reggiani}, {Todorov},
  \& {Snik}}]{2020MNRAS.492..431B}
{Bohn}, A.~J., {Kenworthy}, M.~A., {Ginski}, C., {et~al.} 2020, \mnras, 492,
  431

\bibitem[{{Bonnefoy} {et~al.}(2014){Bonnefoy}, {Marleau}, {Galicher}, {Beust},
  {Lagrange}, {Baudino}, {Chauvin}, {Borgniet}, {Meunier}, {Rameau},
  {Boccaletti}, {Cumming}, {Helling}, {Homeier}, {Allard}, \&
  {Delorme}}]{2014A&A...567L...9B}
{Bonnefoy}, M., {Marleau}, G.~D., {Galicher}, R., {et~al.} 2014, \aap, 567, L9

\bibitem[{{Bonnefoy} {et~al.}(2016){Bonnefoy}, {Zurlo}, {Baudino}, {Lucas},
  {Mesa}, {Maire}, {Vigan}, {Galicher}, {Homeier}, {Marocco}, {Gratton},
  {Chauvin}, {Allard}, {Desidera}, {Kasper}, {Moutou}, {Lagrange}, {Antichi},
  {Baruffolo}, {Baudrand }, {Beuzit}, {Boccaletti}, {Cantalloube}, {Carbillet},
  {Charton}, {Claudi}, {Costille}, {Dohlen}, {Dominik}, {Fantinel},
  {Feautrier}, {Feldt}, {Fusco}, {Gigan}, {Girard}, {Gluck}, {Gry}, {Henning},
  {Janson}, {Langlois}, {Madec}, {Magnard}, {Maurel}, {Mawet}, {Meyer},
  {Milli}, {Moeller-Nilsson}, {Mouillet}, {Pavlov}, {Perret}, {Pujet}, {Quanz},
  {Rochat}, {Rousset}, {Roux}, {Salasnich}, {Salter}, {Sauvage}, {Schmid},
  {Sevin}, {Soenke}, {Stadler}, {Turatto}, {Udry}, {Vakili}, {Wahhaj}, \&
  {Wildi}}]{2016A&A...587A..58B}
{Bonnefoy}, M., {Zurlo}, A., {Baudino}, J.~L., {et~al.} 2016, \aap, 587, A58

\bibitem[{{Booth} {et~al.}(2016){Booth}, {Jord{\'a}n}, {Casassus}, {Hales},
  {Dent}, {Faramaz}, {Matr{\`a}}, {Barkats}, {Brahm}, \&
  {Cuadra}}]{2016MNRAS.460L..10B}
{Booth}, M., {Jord{\'a}n}, A., {Casassus}, S., {et~al.} 2016, \mnras, 460, L10

\bibitem[{{Buenzli} {et~al.}(2015){Buenzli}, {Saumon}, {Marley}, {Apai},
  {Radigan}, {Bedin}, {Reid}, \& {Morley}}]{2015ApJ...798..127B}
{Buenzli}, E., {Saumon}, D., {Marley}, M.~S., {et~al.} 2015, \apj, 798, 127

\bibitem[{{Cantalloube} {et~al.}(2015){Cantalloube}, {Mouillet}, {Mugnier},
  {Milli}, {Absil}, {Gomez Gonzalez}, {Chauvin}, {Beuzit}, \&
  {Cornia}}]{2015A&A...582A..89C}
{Cantalloube}, F., {Mouillet}, D., {Mugnier}, L.~M., {et~al.} 2015, \aap, 582,
  A89

\bibitem[{{Chatterjee} {et~al.}(2008){Chatterjee}, {Ford}, {Matsumura}, \&
  {Rasio}}]{2008ApJ...686..580C}
{Chatterjee}, S., {Ford}, E.~B., {Matsumura}, S., \& {Rasio}, F.~A. 2008, \apj,
  686, 580

\bibitem[{{Chauvin} {et~al.}(2017){Chauvin}, {Desidera}, {Lagrange}, {Vigan},
  {Gratton}, {Langlois}, {Bonnefoy}, {Beuzit}, {Feldt}, {Mouillet}, {Meyer},
  {Cheetham}, {Biller}, {Boccaletti}, {D'Orazi}, {Galicher}, {Hagelberg},
  {Maire}, {Mesa}, {Olofsson}, {Samland}, {Schmidt}, {Sissa}, {Bonavita},
  {Charnay}, {Cudel}, {Daemgen}, {Delorme}, {Janin-Potiron}, {Janson},
  {Keppler}, {Le Coroller}, {Ligi}, {Marleau}, {Messina}, {Molli{\`e}re},
  {Mordasini}, {M{\"u}ller}, {Peretti}, {Perrot}, {Rodet}, {Rouan}, {Zurlo},
  {Dominik}, {Henning}, {Menard}, {Schmid}, {Turatto}, {Udry}, {Vakili}, {Abe},
  {Antichi}, {Baruffolo}, {Baudoz}, {Baudrand}, {Blanchard}, {Bazzon}, {Buey},
  {Carbillet}, {Carle}, {Charton}, {Cascone}, {Claudi}, {Costille}, {Deboulbe},
  {De Caprio}, {Dohlen}, {Fantinel}, {Feautrier}, {Fusco}, {Gigan}, {Giro},
  {Gisler}, {Gluck}, {Hubin}, {Hugot}, {Jaquet}, {Kasper}, {Madec}, {Magnard},
  {Martinez}, {Maurel}, {Le Mignant}, {M{\"o}ller-Nilsson}, {Llored}, {Moulin},
  {Orign{\'e}}, {Pavlov}, {Perret}, {Petit}, {Pragt}, {Puget}, {Rabou},
  {Ramos}, {Rigal}, {Rochat}, {Roelfsema}, {Rousset}, {Roux}, {Salasnich},
  {Sauvage}, {Sevin}, {Soenke}, {Stadler}, {Suarez}, {Weber}, {Wildi},
  {Antoniucci}, {Augereau}, {Baudino}, {Brandner}, {Engler}, {Girard}, {Gry},
  {Kral}, {Kopytova}, {Lagadec}, {Milli}, {Moutou}, {Schlieder},
  {Szul{\'a}gyi}, {Thalmann}, \& {Wahhaj}}]{2017A&A...605L...9C}
{Chauvin}, G., {Desidera}, S., {Lagrange}, A.~M., {et~al.} 2017, \aap, 605, L9

\bibitem[{{Choquet} {et~al.}(2017){Choquet}, {Milli}, {Wahhaj}, {Soummer},
  {Roberge}, {Augereau}, {Booth}, {Absil}, {Boccaletti}, {Chen}, {Debes}, {del
  Burgo}, {Dent}, {Ertel}, {Girard}, {Gofas-Salas}, {Golimowski}, {G{\'o}mez
  Gonz{\'a}lez}, {Hagan}, {Hibon}, {Hines}, {Kennedy}, {Lagrange}, {Matr{\`a}},
  {Mawet}, {Mouillet}, {N'Diaye}, {Perrin}, {Pinte}, {Pueyo}, {Rajan},
  {Schneider}, {Wolff}, \& {Wyatt}}]{2017ApJ...834L..12C}
{Choquet}, {\'E}., {Milli}, J., {Wahhaj}, Z., {et~al.} 2017, \apjl, 834, L12

\bibitem[{{Choquet} {et~al.}(2016){Choquet}, {Perrin}, {Chen}, {Soummer},
  {Pueyo}, {Hagan}, {Gofas-Salas}, {Rajan}, {Golimowski}, {Hines}, {Schneider},
  {Mazoyer}, {Augereau}, {Debes}, {Stark}, {Wolff}, {N'Diaye}, \&
  {Hsiao}}]{2016ApJ...817L...2C}
{Choquet}, {\'E}., {Perrin}, M.~D., {Chen}, C.~H., {et~al.} 2016, \apjl, 817,
  L2

\bibitem[{{Claudi} {et~al.}(2008){Claudi}, {Turatto}, {Gratton}, {Antichi},
  {Bonavita}, {Bruno}, {Cascone}, {De Caprio}, {Desidera}, {Giro}, {Mesa},
  {Scuderi}, {Dohlen}, {Beuzit}, \& {Puget}}]{2008SPIE.7014E..3EC}
{Claudi}, R.~U., {Turatto}, M., {Gratton}, R.~G., {et~al.} 2008, in Society of
  Photo-Optical Instrumentation Engineers (SPIE) Conference Series, Vol. 7014,
  \procspie, 70143E

\bibitem[{{Cowley} {et~al.}(1969){Cowley}, {Cowley}, {Jaschek}, \&
  {Jaschek}}]{1969AJ.....74..375C}
{Cowley}, A., {Cowley}, C., {Jaschek}, M., \& {Jaschek}, C. 1969, \aj, 74, 375

\bibitem[{{Crida}(2009)}]{2009sf2a.conf..313C}
{Crida}, A. 2009, in SF2A-2009: Proceedings of the Annual meeting of the French
  Society of Astronomy and Astrophysics, ed. M.~{Heydari-Malayeri},
  C.~{Reyl'E}, \& R.~{Samadi}, 313

\bibitem[{{Currie} {et~al.}(2011){Currie}, {Burrows}, {Itoh}, {Matsumura},
  {Fukagawa}, {Apai}, {Madhusudhan}, {Hinz}, {Rodigas}, {Kasper}, {Pyo}, \&
  {Ogino}}]{2011ApJ...729..128C}
{Currie}, T., {Burrows}, A., {Itoh}, Y., {et~al.} 2011, \apj, 729, 128

\bibitem[{{Delorme} {et~al.}(2017){Delorme}, {Meunier}, {Albert}, {Lagadec},
  {Le Coroller}, {Galicher}, {Mouillet}, {Boccaletti}, {Mesa}, {Meunier},
  {Beuzit}, {Lagrange}, {Chauvin}, {Sapone}, {Langlois}, {Maire},
  {Montarg{\`e}s}, {Gratton}, {Vigan}, \& {Surace}}]{2017sf2a.conf..347D}
{Delorme}, P., {Meunier}, N., {Albert}, D., {et~al.} 2017, in SF2A-2017:
  Proceedings of the Annual meeting of the French Society of Astronomy and
  Astrophysics, ed. C.~{Reyl{\'e}}, P.~{Di Matteo}, F.~{Herpin}, E.~{Lagadec},
  A.~{Lan{\c{c}}on}, Z.~{Meliani}, \& F.~{Royer}, Di

\bibitem[{{Dohlen} {et~al.}(2008){Dohlen}, {Langlois}, {Saisse}, {Hill},
  {Origne}, {Jacquet}, {Fabron}, {Blanc}, {Llored}, {Carle}, {Moutou}, {Vigan},
  {Boccaletti}, {Carbillet}, {Mouillet}, \& {Beuzit}}]{2008SPIE.7014E..3LD}
{Dohlen}, K., {Langlois}, M., {Saisse}, M., {et~al.} 2008, in Society of
  Photo-Optical Instrumentation Engineers (SPIE) Conference Series, Vol. 7014,
  \procspie, 70143L

\bibitem[{{Dong} \& {Zhu}(2013)}]{2013ApJ...778...53D}
{Dong}, S. \& {Zhu}, Z. 2013, \apj, 778, 53

\bibitem[{{Fabrycky} \& {Murray-Clay}(2010)}]{2010ApJ...710.1408F}
{Fabrycky}, D.~C. \& {Murray-Clay}, R.~A. 2010, \apj, 710, 1408

\bibitem[{{Fernandes} {et~al.}(2019){Fernandes}, {Mulders}, {Pascucci},
  {Mordasini}, \& {Emsenhuber}}]{2019ApJ...874...81F}
{Fernandes}, R.~B., {Mulders}, G.~D., {Pascucci}, I., {Mordasini}, C., \&
  {Emsenhuber}, A. 2019, \apj, 874, 81

\bibitem[{{Fusco} {et~al.}(2005){Fusco}, {Petit}, {Rousset}, {Conan}, \&
  {Beuzit}}]{2005OptL...30.1255F}
{Fusco}, T., {Petit}, C., {Rousset}, G., {Conan}, J.~M., \& {Beuzit}, J.~L.
  2005, Optics Letters, 30, 1255

\bibitem[{{Fusco} {et~al.}(2006){Fusco}, {Rousset}, {Sauvage}, {Petit},
  {Beuzit}, {Dohlen}, {Mouillet}, {Charton}, {Nicolle}, {Kasper}, {Baudoz}, \&
  {Puget}}]{2006OExpr..14.7515F}
{Fusco}, T., {Rousset}, G., {Sauvage}, J.~F., {et~al.} 2006, Optics Express,
  14, 7515

\bibitem[{{Gagn{\'e}} {et~al.}(2014){Gagn{\'e}}, {Lafreni{\`e}re}, {Doyon},
  {Malo}, \& {Artigau}}]{2014ApJ...783..121G}
{Gagn{\'e}}, J., {Lafreni{\`e}re}, D., {Doyon}, R., {Malo}, L., \& {Artigau},
  {\'E}. 2014, \apj, 783, 121

\bibitem[{{Gaia Collaboration}(2018)}]{2018yCat.1345....0G}
{Gaia Collaboration}. 2018, VizieR Online Data Catalog, I/345

\bibitem[{{Galicher} {et~al.}(2011){Galicher}, {Marois}, {Macintosh}, {Barman},
  \& {Konopacky}}]{2011ApJ...739L..41G}
{Galicher}, R., {Marois}, C., {Macintosh}, B., {Barman}, T., \& {Konopacky}, Q.
  2011, \apjl, 739, L41

\bibitem[{{Gauza} {et~al.}(2015){Gauza}, {B{\'e}jar}, {P{\'e}rez-Garrido},
  {Zapatero Osorio}, {Lodieu}, {Rebolo}, {Pall{\'e}}, \&
  {Nowak}}]{2015ApJ...804...96G}
{Gauza}, B., {B{\'e}jar}, V. J.~S., {P{\'e}rez-Garrido}, A., {et~al.} 2015,
  \apj, 804, 96

\bibitem[{{Geiler} {et~al.}(2019){Geiler}, {Krivov}, {Booth}, \&
  {L{\"o}hne}}]{2019MNRAS.483..332G}
{Geiler}, F., {Krivov}, A.~V., {Booth}, M., \& {L{\"o}hne}, T. 2019, \mnras,
  483, 332

\bibitem[{{Go{\'z}dziewski} \& {Migaszewski}(2009)}]{2009MNRAS.397L..16G}
{Go{\'z}dziewski}, K. \& {Migaszewski}, C. 2009, \mnras, 397, L16

\bibitem[{{Go{\'z}dziewski} \& {Migaszewski}(2014)}]{2014MNRAS.440.3140G}
{Go{\'z}dziewski}, K. \& {Migaszewski}, C. 2014, \mnras, 440, 3140

\bibitem[{{Go{\'z}dziewski} \& {Migaszewski}(2018)}]{2018ApJS..238....6G}
{Go{\'z}dziewski}, K. \& {Migaszewski}, C. 2018, \apjs, 238, 6

\bibitem[{{Haffert} {et~al.}(2019){Haffert}, {Bohn}, {de Boer}, {Snellen},
  {Brinchmann}, {Girard}, {Keller}, \& {Bacon}}]{2019NatAs...3..749H}
{Haffert}, S.~Y., {Bohn}, A.~J., {de Boer}, J., {et~al.} 2019, Nature
  Astronomy, 3, 749

\bibitem[{{Hinz} {et~al.}(2010){Hinz}, {Rodigas}, {Kenworthy}, {Sivanandam},
  {Heinze}, {Mamajek}, \& {Meyer}}]{2010ApJ...716..417H}
{Hinz}, P.~M., {Rodigas}, T.~J., {Kenworthy}, M.~A., {et~al.} 2010, \apj, 716,
  417

\bibitem[{{Howard} {et~al.}(2010){Howard}, {Marcy}, {Johnson}, {Fischer},
  {Wright}, {Isaacson}, {Valenti}, {Anderson}, {Lin}, \&
  {Ida}}]{2010Sci...330..653H}
{Howard}, A.~W., {Marcy}, G.~W., {Johnson}, J.~A., {et~al.} 2010, Science, 330,
  653

\bibitem[{{Hughes} {et~al.}(2011){Hughes}, {Wilner}, {Andrews}, {Williams},
  {Su}, {Murray-Clay}, \& {Qi}}]{2011ApJ...740...38H}
{Hughes}, A.~M., {Wilner}, D.~J., {Andrews}, S.~M., {et~al.} 2011, \apj, 740,
  38

\bibitem[{{Hugot} {et~al.}(2012){Hugot}, {Ferrari}, {El Hadi}, {Costille},
  {Dohlen}, {Rabou}, {Puget}, \& {Beuzit}}]{2012A&A...538A.139H}
{Hugot}, E., {Ferrari}, M., {El Hadi}, K., {et~al.} 2012, \aap, 538, A139

\bibitem[{{Ingraham} {et~al.}(2014){Ingraham}, {Marley}, {Saumon}, {Marois},
  {Macintosh}, {Barman}, {Bauman}, {Burrows}, {Chilcote}, {De Rosa}, {Dillon},
  {Doyon}, {Dunn}, {Erikson}, {Fitzgerald}, {Gavel}, {Goodsell}, {Graham},
  {Hartung}, {Hibon}, {Kalas}, {Konopacky}, {Larkin}, {Maire}, {Marchis},
  {McBride}, {Millar-Blanchaer}, {Morzinski}, {Norton}, {Oppenheimer},
  {Palmer}, {Patience}, {Perrin}, {Poyneer}, {Pueyo}, {Rantakyr{\"o}},
  {Sadakuni}, {Saddlemyer}, {Savransky}, {Soummer}, {Sivaramakrishnan}, {Song},
  {Thomas}, {Wallace}, {Wiktorowicz}, \& {Wolff}}]{2014ApJ...794L..15I}
{Ingraham}, P., {Marley}, M.~S., {Saumon}, D., {et~al.} 2014, \apjl, 794, L15

\bibitem[{{Janson} {et~al.}(2010){Janson}, {Bergfors}, {Goto}, {Brandner}, \&
  {Lafreni{\`e}re}}]{2010ApJ...710L..35J}
{Janson}, M., {Bergfors}, C., {Goto}, M., {Brandner}, W., \& {Lafreni{\`e}re},
  D. 2010, \apjl, 710, L35

\bibitem[{{Keppler} {et~al.}(2018){Keppler}, {Benisty}, {M{\"u}ller},
  {Henning}, {van Boekel}, {Cantalloube}, {Ginski}, {van Holstein}, {Maire},
  {Pohl}, {Samland }, {Avenhaus}, {Baudino}, {Boccaletti}, {de Boer},
  {Bonnefoy}, {Chauvin}, {Desidera}, {Langlois}, {Lazzoni}, {Marleau},
  {Mordasini}, {Pawellek}, {Stolker}, {Vigan}, {Zurlo}, {Birnstiel},
  {Brandner}, {Feldt}, {Flock}, {Girard}, {Gratton}, {Hagelberg}, {Isella},
  {Janson}, {Juhasz}, {Kemmer}, {Kral}, {Lagrange}, {Launhardt}, {Matter},
  {M{\'e}nard}, {Milli}, {Molli{\`e}re}, {Olofsson}, {P{\'e}rez}, {Pinilla},
  {Pinte}, {Quanz}, {Schmidt}, {Udry}, {Wahhaj}, {Williams}, {Buenzli},
  {Cudel}, {Dominik}, {Galicher}, {Kasper}, {Lannier}, {Mesa}, {Mouillet},
  {Peretti}, {Perrot}, {Salter}, {Sissa}, {Wildi}, {Abe}, {Antichi},
  {Augereau}, {Baruffolo}, {Baudoz}, {Bazzon}, {Beuzit}, {Blanchard}, {Brems},
  {Buey}, {De Caprio}, {Carbillet}, {Carle}, {Cascone}, {Cheetham}, {Claudi},
  {Costille}, {Delboulb{\'e}}, {Dohlen}, {Fantinel}, {Feautrier}, {Fusco},
  {Giro}, {Gluck}, {Gry}, {Hubin}, {Hugot}, {Jaquet}, {Le Mignant}, {Llored},
  {Madec}, {Magnard}, {Martinez}, {Maurel}, {Meyer}, {M{\"o}ller-Nilsson},
  {Moulin}, {Mugnier}, {Orign{\'e}}, {Pavlov}, {Perret}, {Petit}, {Pragt},
  {Puget}, {Rabou}, {Ramos}, {Rigal}, {Rochat}, {Roelfsema}, {Rousset}, {Roux},
  {Salasnich}, {Sauvage}, {Sevin}, {Soenke}, {Stadler}, {Suarez}, {Turatto}, \&
  {Weber}}]{2018A&A...617A..44K}
{Keppler}, M., {Benisty}, M., {M{\"u}ller}, A., {et~al.} 2018, \aap, 617, A44

\bibitem[{{Konopacky} {et~al.}(2013){Konopacky}, {Barman}, {Macintosh}, \&
  {Marois}}]{2013Sci...339.1398K}
{Konopacky}, Q.~M., {Barman}, T.~S., {Macintosh}, B.~A., \& {Marois}, C. 2013,
  Science, 339, 1398

\bibitem[{{Konopacky} {et~al.}(2016){Konopacky}, {Marois}, {Macintosh},
  {Galicher}, {Barman}, {Metchev}, \& {Zuckerman}}]{2016AJ....152...28K}
{Konopacky}, Q.~M., {Marois}, C., {Macintosh}, B.~A., {et~al.} 2016, \aj, 152,
  28

\bibitem[{{Lafreni{\`e}re} {et~al.}(2007){Lafreni{\`e}re}, {Marois}, {Doyon},
  {Nadeau}, \& {Artigau}}]{2007ApJ...660..770L}
{Lafreni{\`e}re}, D., {Marois}, C., {Doyon}, R., {Nadeau}, D., \& {Artigau},
  {\'E}. 2007, \apj, 660, 770

\bibitem[{{Lagrange} {et~al.}(2012){Lagrange}, {Boccaletti}, {Milli},
  {Chauvin}, {Bonnefoy}, {Mouillet}, {Augereau}, {Girard}, {Lacour}, \&
  {Apai}}]{2012A&A...542A..40L}
{Lagrange}, A.~M., {Boccaletti}, A., {Milli}, J., {et~al.} 2012, \aap, 542, A40

\bibitem[{{Lagrange} {et~al.}(2009){Lagrange}, {Gratadour}, {Chauvin}, {Fusco},
  {Ehrenreich}, {Mouillet}, {Rousset}, {Rouan}, {Allard}, {Gendron}, {Charton},
  {Mugnier}, {Rabou}, {Montri}, \& {Lacombe}}]{2009A&A...493L..21L}
{Lagrange}, A.~M., {Gratadour}, D., {Chauvin}, G., {et~al.} 2009, \aap, 493,
  L21

\bibitem[{{Liu}(2004)}]{2004Sci...305.1442L}
{Liu}, M.~C. 2004, Science, 305, 1442

\bibitem[{{Looper} {et~al.}(2008){Looper}, {Kirkpatrick}, {Cutri}, {Barman},
  {Burgasser}, {Cushing}, {Roellig}, {McGovern}, {McLean}, {Rice}, {Swift}, \&
  {Schurr}}]{2008ApJ...686..528L}
{Looper}, D.~L., {Kirkpatrick}, J.~D., {Cutri}, R.~M., {et~al.} 2008, \apj,
  686, 528

\bibitem[{{Macintosh} {et~al.}(2015){Macintosh}, {Graham}, {Barman}, {De Rosa},
  {Konopacky}, {Marley}, {Marois}, {Nielsen}, {Pueyo}, {Rajan}, {Rameau},
  {Saumon}, {Wang}, {Patience}, {Ammons}, {Arriaga}, {Artigau}, {Beckwith},
  {Brewster}, {Bruzzone}, {Bulger}, {Burningham}, {Burrows}, {Chen}, {Chiang},
  {Chilcote}, {Dawson}, {Dong}, {Doyon}, {Draper}, {Duch{\^e}ne}, {Esposito},
  {Fabrycky}, {Fitzgerald}, {Follette}, {Fortney}, {Gerard}, {Goodsell},
  {Greenbaum}, {Hibon}, {Hinkley}, {Cotten}, {Hung}, {Ingraham},
  {Johnson-Groh}, {Kalas}, {Lafreniere}, {Larkin}, {Lee}, {Line}, {Long},
  {Maire}, {Marchis}, {Matthews}, {Max}, {Metchev}, {Millar-Blanchaer},
  {Mittal}, {Morley}, {Morzinski}, {Murray-Clay}, {Oppenheimer}, {Palmer},
  {Patel}, {Perrin}, {Poyneer}, {Rafikov}, {Rantakyr{\"o}}, {Rice}, {Rojo},
  {Rudy}, {Ruffio}, {Ruiz}, {Sadakuni}, {Saddlemyer}, {Salama}, {Savransky},
  {Schneider}, {Sivaramakrishnan}, {Song}, {Soummer}, {Thomas}, {Vasisht},
  {Wallace}, {Ward-Duong}, {Wiktorowicz}, {Wolff}, \&
  {Zuckerman}}]{2015Sci...350...64M}
{Macintosh}, B., {Graham}, J.~R., {Barman}, T., {et~al.} 2015, Science, 350, 64

\bibitem[{{Madhusudhan}(2019)}]{2019ARA&A..57..617M}
{Madhusudhan}, N. 2019, \araa, 57, 617

\bibitem[{{Madhusudhan} {et~al.}(2011){Madhusudhan}, {Burrows}, \&
  {Currie}}]{2011ApJ...737...34M}
{Madhusudhan}, N., {Burrows}, A., \& {Currie}, T. 2011, \apj, 737, 34

\bibitem[{{Maire} {et~al.}(2016){Maire}, {Bonnefoy}, {Ginski}, {Vigan},
  {Messina}, {Mesa}, {Galicher}, {Gratton}, {Desidera}, {Kopytova}, {Millward},
  {Thalmann}, {Claudi}, {Ehrenreich}, {Zurlo}, {Chauvin}, {Antichi},
  {Baruffolo}, {Bazzon}, {Beuzit}, {Blanchard}, {Boccaletti}, {de Boer},
  {Carle}, {Cascone}, {Costille}, {De Caprio}, {Delboulb{\'e}}, {Dohlen},
  {Dominik}, {Feldt}, {Fusco}, {Girard}, {Giro}, {Gisler}, {Gluck}, {Gry},
  {Henning}, {Hubin}, {Hugot}, {Jaquet}, {Kasper}, {Lagrange}, {Langlois}, {Le
  Mignant}, {Llored}, {Madec}, {Martinez}, {Mawet}, {Milli},
  {M{\"o}ller-Nilsson}, {Mouillet}, {Moulin}, {Moutou}, {Orign{\'e}}, {Pavlov},
  {Petit}, {Pragt}, {Puget}, {Ramos}, {Rochat}, {Roelfsema}, {Salasnich},
  {Sauvage}, {Schmid}, {Turatto}, {Udry}, {Vakili}, {Wahhaj}, {Weber}, \&
  {Wildi}}]{2016A&A...587A..56M}
{Maire}, A.~L., {Bonnefoy}, M., {Ginski}, C., {et~al.} 2016, \aap, 587, A56

\bibitem[{{Marleau} \& {Cumming}(2014)}]{2014MNRAS.437.1378M}
{Marleau}, G.~D. \& {Cumming}, A. 2014, \mnras, 437, 1378

\bibitem[{{Marley} {et~al.}(2012){Marley}, {Saumon}, {Cushing}, {Ackerman},
  {Fortney}, \& {Freedman}}]{2012ApJ...754..135M}
{Marley}, M.~S., {Saumon}, D., {Cushing}, M., {et~al.} 2012, \apj, 754, 135

\bibitem[{{Marois} {et~al.}(2006){Marois}, {Lafreni{\`e}re}, {Doyon},
  {Macintosh}, \& {Nadeau}}]{2006ApJ...641..556M}
{Marois}, C., {Lafreni{\`e}re}, D., {Doyon}, R., {Macintosh}, B., \& {Nadeau},
  D. 2006, \apj, 641, 556

\bibitem[{{Marois} {et~al.}(2008){Marois}, {Macintosh}, {Barman}, {Zuckerman},
  {Song}, {Patience}, {Lafreni{\`e}re}, \& {Doyon}}]{2008Sci...322.1348M}
{Marois}, C., {Macintosh}, B., {Barman}, T., {et~al.} 2008, Science, 322, 1348

\bibitem[{{Marois} {et~al.}(2010){Marois}, {Zuckerman}, {Konopacky},
  {Macintosh}, \& {Barman}}]{2010Natur.468.1080M}
{Marois}, C., {Zuckerman}, B., {Konopacky}, Q.~M., {Macintosh}, B., \&
  {Barman}, T. 2010, \nat, 468, 1080

\bibitem[{{Marshall} {et~al.}(2010){Marshall}, {Horner}, \&
  {Carter}}]{2010IJAsB...9..259M}
{Marshall}, J., {Horner}, J., \& {Carter}, A. 2010, International Journal of
  Astrobiology, 9, 259

\bibitem[{Matthews {et~al.}(2013)Matthews, Kennedy, Sibthorpe, Booth, Wyatt,
  Broekhoven-Fiene, Macintosh, \& Marois}]{Matthews_2013}
Matthews, B., Kennedy, G., Sibthorpe, B., {et~al.} 2013, The Astrophysical
  Journal, 780, 97

\bibitem[{{Matthews} {et~al.}(2014){Matthews}, {Kennedy}, {Sibthorpe}, {Booth},
  {Wyatt}, {Broekhoven-Fiene}, {Macintosh}, \& {Marois}}]{2014ApJ...780...97M}
{Matthews}, B., {Kennedy}, G., {Sibthorpe}, B., {et~al.} 2014, \apj, 780, 97

\bibitem[{{Mayor} {et~al.}(2011){Mayor}, {Marmier}, {Lovis}, {Udry},
  {S{\'e}gransan}, {Pepe}, {Benz}, {Bertaux}, {Bouchy}, {Dumusque}, {Lo Curto},
  {Mordasini}, {Queloz}, \& {Santos}}]{2011arXiv1109.2497M}
{Mayor}, M., {Marmier}, M., {Lovis}, C., {et~al.} 2011, arXiv e-prints,
  arXiv:1109.2497

\bibitem[{{Milli} {et~al.}(2018){Milli}, {Kasper}, {Bourget}, {Pannetier},
  {Mouillet}, {Sauvage}, {Reyes}, {Fusco}, {Cantalloube}, {Tristam}, {Wahhaj},
  {Beuzit}, {Girard}, {Mawet}, {Telle}, {Vigan}, \&
  {N'Diaye}}]{2018SPIE10703E..2AM}
{Milli}, J., {Kasper}, M., {Bourget}, P., {et~al.} 2018, in Society of
  Photo-Optical Instrumentation Engineers (SPIE) Conference Series, Vol. 10703,
  \procspie, 107032A

\bibitem[{{Milli} {et~al.}(2012){Milli}, {Mouillet}, {Lagrange}, {Boccaletti},
  {Mawet}, {Chauvin}, \& {Bonnefoy}}]{2012A&A...545A.111M}
{Milli}, J., {Mouillet}, D., {Lagrange}, A.~M., {et~al.} 2012, \aap, 545, A111

\bibitem[{{Mo{\'o}r} {et~al.}(2006){Mo{\'o}r}, {{\'A}brah{\'a}m}, {Derekas},
  {Kiss}, {Kiss}, {Apai}, {Grady}, \& {Henning}}]{2006ApJ...644..525M}
{Mo{\'o}r}, A., {{\'A}brah{\'a}m}, P., {Derekas}, A., {et~al.} 2006, \apj, 644,
  525

\bibitem[{{Mo{\'o}r} {et~al.}(2015){Mo{\'o}r}, {K{\'o}sp{\'a}l},
  {{\'A}brah{\'a}m}, {Apai}, {Balog}, {Grady}, {Henning}, {Juh{\'a}sz}, {Kiss},
  {Krivov}, {Pawellek}, \& {Szab{\'o}}}]{2015MNRAS.447..577M}
{Mo{\'o}r}, A., {K{\'o}sp{\'a}l}, {\'A}., {{\'A}brah{\'a}m}, P., {et~al.} 2015,
  \mnras, 447, 577

\bibitem[{{Morley} {et~al.}(2012){Morley}, {Fortney}, {Marley}, {Visscher},
  {Saumon}, \& {Leggett}}]{2012ApJ...756..172M}
{Morley}, C.~V., {Fortney}, J.~J., {Marley}, M.~S., {et~al.} 2012, \apj, 756,
  172

\bibitem[{{Moro-Mart{\'\i}n} {et~al.}(2010){Moro-Mart{\'\i}n}, {Malhotra},
  {Bryden}, {Rieke}, {Su}, {Beichman}, \& {Lawler}}]{2010ApJ...717.1123M}
{Moro-Mart{\'\i}n}, A., {Malhotra}, R., {Bryden}, G., {et~al.} 2010, \apj, 717,
  1123

\bibitem[{{M{\"u}ller} {et~al.}(2018){M{\"u}ller}, {Keppler}, {Henning},
  {Samland}, {Chauvin}, {Beust}, {Maire}, {Molaverdikhani}, {van Boekel},
  {Benisty}, {Boccaletti}, {Bonnefoy}, {Cantalloube}, {Charnay}, {Baudino},
  {Gennaro}, {Long}, {Cheetham}, {Desidera}, {Feldt}, {Fusco}, {Girard},
  {Gratton}, {Hagelberg}, {Janson}, {Lagrange}, {Langlois}, {Lazzoni}, {Ligi},
  {M{\'e}nard}, {Mesa}, {Meyer}, {Molli{\`e}re}, {Mordasini}, {Moulin},
  {Pavlov}, {Pawellek}, {Quanz}, {Ramos}, {Rouan}, {Sissa}, {Stadler}, {Vigan},
  {Wahhaj}, {Weber}, \& {Zurlo}}]{2018A&A...617L...2M}
{M{\"u}ller}, A., {Keppler}, M., {Henning}, T., {et~al.} 2018, \aap, 617, L2

\bibitem[{{Nielsen} {et~al.}(2019){Nielsen}, {De Rosa}, {Macintosh}, {Wang},
  {Ruffio}, {Chiang}, {Marley}, {Saumon}, {Savransky}, {Ammons}, {Bailey},
  {Barman}, {Blain}, {Bulger}, {Burrows}, {Chilcote}, {Cotten}, {Czekala},
  {Doyon}, {Duch{\^e}ne}, {Esposito}, {Fabrycky}, {Fitzgerald}, {Follette},
  {Fortney}, {Gerard}, {Goodsell}, {Graham}, {Greenbaum}, {Hibon}, {Hinkley},
  {Hirsch}, {Hom}, {Hung}, {Dawson}, {Ingraham}, {Kalas}, {Konopacky},
  {Larkin}, {Lee}, {Lin}, {Maire}, {Marchis}, {Marois}, {Metchev},
  {Millar-Blanchaer}, {Morzinski}, {Oppenheimer}, {Palmer}, {Patience},
  {Perrin}, {Poyneer}, {Pueyo}, {Rafikov}, {Rajan}, {Rameau}, {Rantakyr{\"o}},
  {Ren}, {Schneider}, {Sivaramakrishnan}, {Song}, {Soummer}, {Tallis},
  {Thomas}, {Ward-Duong}, \& {Wolff}}]{2019AJ....158...13N}
{Nielsen}, E.~L., {De Rosa}, R.~J., {Macintosh}, B., {et~al.} 2019, \aj, 158,
  13

\bibitem[{{Oppenheimer} {et~al.}(2013){Oppenheimer}, {Baranec}, {Beichman},
  {Brenner}, {Burruss}, {Cady}, {Crepp}, {Dekany}, {Fergus}, {Hale},
  {Hillenbrand}, {Hinkley}, {Hogg}, {King}, {Ligon}, {Lockhart}, {Nilsson},
  {Parry}, {Pueyo}, {Rice}, {Roberts}, {Roberts}, {Shao}, {Sivaramakrishnan},
  {Soummer}, {Truong}, {Vasisht}, {Veicht}, {Vescelus}, {Wallace}, {Zhai}, \&
  {Zimmerman}}]{2013ApJ...768...24O}
{Oppenheimer}, B.~R., {Baranec}, C., {Beichman}, C., {et~al.} 2013, \apj, 768,
  24

\bibitem[{{Petit} {et~al.}(2012){Petit}, {Sauvage}, {Sevin}, {Costille},
  {Fusco}, {Baudoz}, {Beuzit}, {Buey}, {Charton}, {Dohlen}, {Feautrier},
  {Fedrigo}, {Gach}, {Hubin}, {Hugot}, {Kasper}, {Mouillet}, {Perret}, {Puget},
  {Sinquin}, {Soenke}, {Suarez}, \& {Wildi}}]{2012SPIE.8447E..1ZP}
{Petit}, C., {Sauvage}, J.~F., {Sevin}, A., {et~al.} 2012, in Society of
  Photo-Optical Instrumentation Engineers (SPIE) Conference Series, Vol. 8447,
  \procspie, 84471Z

\bibitem[{{Pueyo} {et~al.}(2015){Pueyo}, {Soummer}, {Hoffmann}, {Oppenheimer},
  {Graham}, {Zimmerman}, {Zhai}, {Wallace}, {Vescelus}, {Veicht}, {Vasisht},
  {Truong}, {Sivaramakrishnan}, {Shao}, {Roberts}, {Roberts}, {Rice}, {Parry},
  {Nilsson}, {Lockhart}, {Ligon}, {King}, {Hinkley}, {Hillenbrand}, {Hale},
  {Dekany}, {Crepp}, {Cady}, {Burruss}, {Brenner}, {Beichman}, \&
  {Baranec}}]{2015ApJ...803...31P}
{Pueyo}, L., {Soummer}, R., {Hoffmann}, J., {et~al.} 2015, \apj, 803, 31

\bibitem[{{Racine} {et~al.}(1999){Racine}, {Walker}, {Nadeau}, {Doyon}, \&
  {Marois}}]{1999PASP..111..587R}
{Racine}, R., {Walker}, G.~A.~H., {Nadeau}, D., {Doyon}, R., \& {Marois}, C.
  1999, \pasp, 111, 587

\bibitem[{{Raymond} {et~al.}(2010){Raymond}, {Armitage}, \&
  {Gorelick}}]{2010ApJ...711..772R}
{Raymond}, S.~N., {Armitage}, P.~J., \& {Gorelick}, N. 2010, \apj, 711, 772

\bibitem[{{Reidemeister} {et~al.}(2009){Reidemeister}, {Krivov}, {Schmidt},
  {Fiedler}, {M{\"u}ller}, {L{\"o}hne}, \&
  {Neuh{\"a}user}}]{2009A&A...503..247R}
{Reidemeister}, M., {Krivov}, A.~V., {Schmidt}, T.~O.~B., {et~al.} 2009, \aap,
  503, 247

\bibitem[{{Ruane} {et~al.}(2019){Ruane}, {Ngo}, {Mawet}, {Absil}, {Choquet},
  {Cook}, {Gomez Gonzalez}, {Huby}, {Matthews}, {Meshkat}, {Reggiani},
  {Serabyn}, {Wallack}, \& {Xuan}}]{2019AJ....157..118R}
{Ruane}, G., {Ngo}, H., {Mawet}, D., {et~al.} 2019, \aj, 157, 118

\bibitem[{Sadakane \& Nishida(1986)}]{Sadakane_1986}
Sadakane, K. \& Nishida, M. 1986, Publications of the Astronomical Society of
  the Pacific, 98, 685

\bibitem[{{Saio}(2019)}]{2019MNRAS.487.2177S}
{Saio}, H. 2019, \mnras, 487, 2177

\bibitem[{{Saio} {et~al.}(2018){Saio}, {Bedding}, {Kurtz}, {Murphy}, {Antoci},
  {Shibahashi}, {Li}, \& {Takata}}]{2018MNRAS.477.2183S}
{Saio}, H., {Bedding}, T.~R., {Kurtz}, D.~W., {et~al.} 2018, \mnras, 477, 2183

\bibitem[{{Sauvage} {et~al.}(2016){Sauvage}, {Fusco}, {Petit}, {Costille},
  {Mouillet}, {Beuzit}, {Dohlen}, {Kasper}, {Suarez}, {Soenke}, {Baruffolo},
  {Salasnich}, {Rochat}, {Fedrigo}, {Baudoz}, {Hugot}, {Sevin}, {Perret},
  {Wildi}, {Downing}, {Feautrier}, {Puget}, {Vigan}, {O'Neal}, {Girard},
  {Mawet}, {Schmid}, \& {Roelfsema}}]{2016JATIS...2b5003S}
{Sauvage}, J.-F., {Fusco}, T., {Petit}, C., {et~al.} 2016, Journal of
  Astronomical Telescopes, Instruments, and Systems, 2, 025003

\bibitem[{{Schmid} {et~al.}(2018){Schmid}, {Bazzon}, {Roelfsema}, {Mouillet},
  {Milli}, {Menard}, {Gisler}, {Hunziker}, {Pragt}, {Dominik}, {Boccaletti},
  {Ginski}, {Abe}, {Antoniucci}, {Avenhaus}, {Baruffolo}, {Baudoz}, {Beuzit},
  {Carbillet}, {Chauvin}, {Claudi}, {Costille}, {Daban}, {de Haan}, {Desidera},
  {Dohlen}, {Downing}, {Elswijk}, {Engler}, {Feldt}, {Fusco}, {Girard},
  {Gratton}, {Hanenburg}, {Henning}, {Hubin}, {Joos}, {Kasper}, {Keller},
  {Langlois}, {Lagadec}, {Martinez}, {Mulder}, {Pavlov}, {Podio}, {Puget},
  {Quanz}, {Rigal}, {Salasnich}, {Sauvage}, {Schuil}, {Siebenmorgen}, {Sissa},
  {Snik}, {Suarez}, {Thalmann}, {Turatto}, {Udry}, {van Duin}, {van Holstein},
  {Vigan}, \& {Wildi}}]{2018A&A...619A...9S}
{Schmid}, H.~M., {Bazzon}, A., {Roelfsema}, R., {et~al.} 2018, \aap, 619, A9

\bibitem[{{Skemer} {et~al.}(2012){Skemer}, {Hinz}, {Esposito}, {Burrows},
  {Leisenring}, {Skrutskie}, {Desidera}, {Mesa}, {Arcidiacono}, {Mannucci},
  {Rodigas}, {Close}, {McCarthy}, {Kulesa}, {Agapito}, {Apai}, {Argomedo},
  {Bailey}, {Boutsia}, {Briguglio}, {Brusa}, {Busoni}, {Claudi}, {Eisner},
  {Fini}, {Follette}, {Garnavich}, {Gratton}, {Guerra}, {Hill}, {Hoffmann},
  {Jones}, {Krejny}, {Males}, {Masciadri}, {Meyer}, {Miller}, {Morzinski},
  {Nelson}, {Pinna}, {Puglisi}, {Quanz}, {Quiros-Pacheco}, {Riccardi},
  {Stefanini}, {Vaitheeswaran}, {Wilson}, \& {Xompero}}]{2012ApJ...753...14S}
{Skemer}, A.~J., {Hinz}, P.~M., {Esposito}, S., {et~al.} 2012, \apj, 753, 14

\bibitem[{{Soummer}(2005)}]{2005ApJ...618L.161S}
{Soummer}, R. 2005, \apjl, 618, L161

\bibitem[{{Soummer} {et~al.}(2012){Soummer}, {Pueyo}, \&
  {Larkin}}]{2012ApJ...755L..28S}
{Soummer}, R., {Pueyo}, L., \& {Larkin}, J. 2012, \apjl, 755, L28

\bibitem[{{Sparks} \& {Ford}(2002)}]{2002ApJ...578..543S}
{Sparks}, W.~B. \& {Ford}, H.~C. 2002, \apj, 578, 543

\bibitem[{{Stephens} {et~al.}(2009){Stephens}, {Leggett}, {Cushing}, {Marley},
  {Saumon}, {Geballe}, {Golimowski}, {Fan}, \& {Noll}}]{2009ApJ...702..154S}
{Stephens}, D.~C., {Leggett}, S.~K., {Cushing}, M.~C., {et~al.} 2009, \apj,
  702, 154

\bibitem[{Su {et~al.}(2009)Su, Rieke, Stapelfeldt, Malhotra, Bryden, Smith,
  Misselt, Moro-Martin, \& Williams}]{Su_2009}
Su, K. Y.~L., Rieke, G.~H., Stapelfeldt, K.~R., {et~al.} 2009, The
  Astrophysical Journal, 705, 314

\bibitem[{{Su} {et~al.}(2009){Su}, {Rieke}, {Stapelfeldt}, {Malhotra},
  {Bryden}, {Smith}, {Misselt}, {Moro-Martin}, \&
  {Williams}}]{2009ApJ...705..314S}
{Su}, K.~Y.~L., {Rieke}, G.~H., {Stapelfeldt}, K.~R., {et~al.} 2009, \apj, 705,
  314

\bibitem[{{Sudol} \& {Haghighipour}(2012)}]{2012ApJ...755...38S}
{Sudol}, J.~J. \& {Haghighipour}, N. 2012, \apj, 755, 38

\bibitem[{{Takata} {et~al.}(2020){Takata}, {Ouazzani}, {Saio}, {Christophe},
  {Ballot}, {Antoci}, {Salmon}, \& {Hijikawa}}]{2020A&A...635A.106T}
{Takata}, M., {Ouazzani}, R.~M., {Saio}, H., {et~al.} 2020, \aap, 635, A106

\bibitem[{{Torres} {et~al.}(2008){Torres}, {Quast}, {Melo}, \&
  {Sterzik}}]{2008hsf2.book..757T}
{Torres}, C.~A.~O., {Quast}, G.~R., {Melo}, C.~H.~F., \& {Sterzik}, M.~F. 2008,
  {Young Nearby Loose Associations}, ed. {Reipurth, B.}, 757

\bibitem[{{Vigan} {et~al.}(2015){Vigan}, {Gry}, {Salter}, {Mesa}, {Homeier},
  {Moutou}, \& {Allard}}]{2015MNRAS.454..129V}
{Vigan}, A., {Gry}, C., {Salter}, G., {et~al.} 2015, \mnras, 454, 129

\bibitem[{{Vigan} {et~al.}(2010){Vigan}, {Moutou}, {Langlois}, {Allard},
  {Boccaletti}, {Carbillet}, {Mouillet}, \& {Smith}}]{2010MNRAS.407...71V}
{Vigan}, A., {Moutou}, C., {Langlois}, M., {et~al.} 2010, \mnras, 407, 71

\bibitem[{{Wahhaj} {et~al.}(2015){Wahhaj}, {Cieza}, {Mawet}, {Yang}, {Canovas},
  {de Boer}, {Casassus}, {M{\'e}nard}, {Schreiber}, {Liu}, {Biller}, {Nielsen},
  \& {Hayward}}]{2015A&A...581A..24W}
{Wahhaj}, Z., {Cieza}, L.~A., {Mawet}, D., {et~al.} 2015, \aap, 581, A24

\bibitem[{{Wahhaj} {et~al.}(2011){Wahhaj}, {Liu}, {Biller}, {Clarke},
  {Nielsen}, {Close}, {Hayward}, {Mamajek}, {Cushing}, {Dupuy}, {Tecza},
  {Thatte}, {Chun}, {Ftaclas}, {Hartung}, {Reid}, {Shkolnik}, {Alencar},
  {Artymowicz}, {Boss}, {de Gouveia Dal Pino}, {Gregorio-Hetem}, {Ida},
  {Kuchner}, {Lin}, \& {Toomey}}]{2011ApJ...729..139W}
{Wahhaj}, Z., {Liu}, M.~C., {Biller}, B.~A., {et~al.} 2011, \apj, 729, 139

\bibitem[{{Wahhaj} {et~al.}(2013){Wahhaj}, {Liu}, {Biller}, {Nielsen}, {Close},
  {Hayward}, \& {Others}}]{2013ApJ...779...80W}
{Wahhaj}, Z., {Liu}, M.~C., {Biller}, B.~A., {et~al.} 2013, \apj, 779, 80

\bibitem[{{Wahhaj} {et~al.}(2016){Wahhaj}, {Milli}, {Kennedy}, {Ertel},
  {Matr{\`a}}, {Boccaletti}, {del Burgo}, {Wyatt}, {Pinte}, {Lagrange},
  {Absil}, {Choquet}, {G{\'o}mez Gonz{\'a}lez}, {Kobayashi}, {Mawet},
  {Mouillet}, {Pueyo}, {Dent}, {Augereau}, \& {Girard}}]{2016A&A...596L...4W}
{Wahhaj}, Z., {Milli}, J., {Kennedy}, G., {et~al.} 2016, \aap, 596, L4

\bibitem[{{Wang} {et~al.}(2018){Wang}, {Graham}, {Dawson}, {Fabrycky}, {De
  Rosa}, {Pueyo}, {Konopacky}, {Macintosh}, {Marois}, {Chiang}, {Ammons},
  {Arriaga}, {Bailey}, {Barman}, {Bulger}, {Chilcote}, {Cotten}, {Doyon},
  {Duch{\^e}ne}, {Esposito}, {Fitzgerald}, {Follette}, {Gerard}, {Goodsell},
  {Greenbaum}, {Hibon}, {Hung}, {Ingraham}, {Kalas}, {Larkin}, {Maire},
  {Marchis}, {Marley}, {Metchev}, {Millar-Blanchaer}, {Nielsen}, {Oppenheimer},
  {Palmer}, {Patience}, {Perrin}, {Poyneer}, {Rajan}, {Rameau},
  {Rantakyr{\"o}}, {Ruffio}, {Savransky}, {Schneider}, {Sivaramakrishnan},
  {Song}, {Soummer}, {Thomas}, {Wallace}, {Ward-Duong}, {Wiktorowicz}, \&
  {Wolff}}]{2018AJ....156..192W}
{Wang}, J.~J., {Graham}, J.~R., {Dawson}, R., {et~al.} 2018, \aj, 156, 192

\bibitem[{{Weinberger} {et~al.}(1999){Weinberger}, {Becklin}, {Schneider},
  {Smith}, {Lowrance}, {Silverstone}, {Zuckerman}, \&
  {Terrile}}]{1999ApJ...525L..53W}
{Weinberger}, A.~J., {Becklin}, E.~E., {Schneider}, G., {et~al.} 1999, \apjl,
  525, L53

\bibitem[{{Wilner} {et~al.}(2018){Wilner}, {MacGregor}, {Andrews}, {Hughes},
  {Matthews}, \& {Su}}]{2018ApJ...855...56W}
{Wilner}, D.~J., {MacGregor}, M.~A., {Andrews}, S.~M., {et~al.} 2018, \apj,
  855, 56

\bibitem[{{Xuan} {et~al.}(2018){Xuan}, {Mawet}, {Ngo}, {Ruane}, {Bailey},
  {Choquet}, {Absil}, {Alvarez}, {Bryan}, {Cook}, {Femen{\'\i}a Castell{\'a}},
  {Gomez Gonzalez}, {Huby}, {Knutson}, {Matthews}, {Ragland}, {Serabyn}, \&
  {Zawol}}]{2018AJ....156..156X}
{Xuan}, W.~J., {Mawet}, D., {Ngo}, H., {et~al.} 2018, \aj, 156, 156

\bibitem[{{Zuckerman} {et~al.}(2011){Zuckerman}, {Rhee}, {Song}, \&
  {Bessell}}]{2011ApJ...732...61Z}
{Zuckerman}, B., {Rhee}, J.~H., {Song}, I., \& {Bessell}, M.~S. 2011, \apj,
  732, 61

\bibitem[{{Zurlo} {et~al.}(2016){Zurlo}, {Vigan}, {Galicher}, {Maire}, {Mesa},
  {Gratton}, {Chauvin}, {Kasper}, {Moutou}, {Bonnefoy}, {Desidera}, {Abe},
  {Apai}, {Baruffolo}, {Baudoz}, {Baudrand}, {Beuzit}, {Blancard},
  {Boccaletti}, {Cantalloube}, {Carle}, {Cascone}, {Charton}, {Claudi},
  {Costille}, {de Caprio}, {Dohlen}, {Dominik}, {Fantinel}, {Feautrier},
  {Feldt}, {Fusco}, {Gigan}, {Girard}, {Gisler}, {Gluck}, {Gry}, {Henning},
  {Hugot}, {Janson}, {Jaquet}, {Lagrange}, {Langlois}, {Llored}, {Madec},
  {Magnard}, {Martinez}, {Maurel}, {Mawet}, {Meyer}, {Milli},
  {Moeller-Nilsson}, {Mouillet}, {Orign{\'e}}, {Pavlov}, {Petit}, {Puget},
  {Quanz}, {Rabou}, {Ramos}, {Rousset}, {Roux}, {Salasnich}, {Salter},
  {Sauvage}, {Schmid}, {Soenke}, {Stadler}, {Suarez}, {Turatto}, {Udry},
  {Vakili}, {Wahhaj}, {Wildi}, \& {Antichi}}]{2016A&A...587A..57Z}
{Zurlo}, A., {Vigan}, A., {Galicher}, R., {et~al.} 2016, \aap, 587, A57

\bibitem[{{Zurlo} {et~al.}(2014){Zurlo}, {Vigan}, {Mesa}, {Gratton}, {Moutou},
  {Langlois}, {Claudi}, {Pueyo}, {Boccaletti}, {Baruffolo}, {Beuzit},
  {Costille}, {Desidera}, {Dohlen}, {Feldt}, {Fusco}, {Henning}, {Kasper},
  {Martinez}, {Moeller-Nilsson}, {Mouillet}, {Pavlov}, {Puget}, {Sauvage},
  {Turatto}, {Udry}, {Vakili}, {Waters}, \& {Wildi}}]{2014A&A...572A..85Z}
{Zurlo}, A., {Vigan}, A., {Mesa}, D., {et~al.} 2014, \aap, 572, A85

\end{thebibliography}

\end{document}